\documentclass{article}

\usepackage{arxiv}

\usepackage[utf8]{inputenc} 
\usepackage[T1]{fontenc}    
\usepackage{hyperref}       
\usepackage{url}            
\usepackage{booktabs}       
\usepackage{amsfonts}       
\usepackage{nicefrac}       
\usepackage{microtype}      
\usepackage{lipsum}		
\usepackage{graphicx}
\usepackage{natbib}
\usepackage{doi}
\usepackage{caption}
\usepackage{subcaption}
\usepackage{algorithm}
\usepackage{algorithmic}
\usepackage{amsmath}

\title{Representing and Modeling Inconsistent, \\Impossible and Incoherent Shapes and Scenes \\with 2D  Non-Conservative Vector Fields \\mapped on 2-Complexes}

\author{ \href{https://orcid.org/0000-0003-3618-4166}{\includegraphics[scale=0.06]{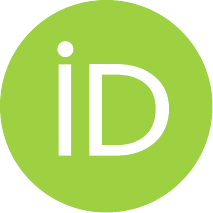}\hspace{1mm}Ergun Akleman}\thanks{Joint with Computer Science and Engineering Department.} \\
	Visual Computing \& Computational Media,\\ Texas A\&M University, College Station, TX, 77831\\
	\texttt{ergun@tamu.edu} \\
       \And
      Youyou Wang\\
Department of Computer Science and Engineering\\ 
Texas A\&M University, College Station, TX, 77831\\
	\texttt{kingyy2010@gmail.com} \\
      \And
	\"Ozg\"ur G\"onen\\
Department of Architecture\\ 
Texas A\&M University, College Station, TX, 77831\\
	\texttt{ozgur.gonen@gmail.com} \\
}

\renewcommand{\shorttitle}{Modeling Inconsistent, Impossible and Incoherent Shapes}

\hypersetup{
pdftitle={A template for the arxiv style},
pdfsubject={q-bio.NC, q-bio.QM},
pdfauthor={David S.~Hippocampus, Elias D.~Striatum},
pdfkeywords={First keyword, Second keyword, More},
}

\begin{document}
\maketitle

\begin{abstract}
In this paper, we present a framework to represent mock 3D objects and scenes, which are not 3D but appear 3D. In our framework, each mock-3D object is represented using 2D non-conservative vector fields and thickness information that are mapped on 2-complexes. Mock-3D scenes are simply scenes consisting of more than one mock-3D object.  
We demonstrated that using this representation, we can dynamically compute a 3D shape using rays emanating from any given point in 3D. These mock-3D objects are view-dependent since their computed shapes depend on the positions of ray centers. Using these dynamically computed shapes, we can compute shadows, reflections, and refractions in real time. 
This representation is mainly useful for 2D artistic applications to model incoherent, inconsistent, and impossible objects. Using this representation, it is possible to obtain expressive depictions with shadows and global illumination effects. The representation can also be used to convert existing 2D artworks into a Mock-3D form that can be interactively re-rendered. 
\end{abstract}  

\section{Introduction}

Despite the significant advances made in 3D computer graphics and shape modeling, according to recent market research 3D Graphics is still only 8\% of the whole graphics market, while 2D graphics markets such as vector, image, and video constitute the rest, i.e. more than 90\%, of the graphics market \cite{hart2008}. Moreover, the 3D modeling market does not grow as rapidly as the 2D painting/editing market. 

There are several usual suspects to explain the reluctance to adapt 3D modeling such that 3D modeling is less intuitive, more expensive, and requires more training than 2D. We think that there exists an additional and important reason. using 3D, it is hard to include all types of expressive depictions that are caused by impossible, inconsistent, and incoherent shapes. 

Although this can be viewed as a problem in the shape modeling community which is mainly focused on 3D, we think that this shortcoming presents an opportunity for the community to explore new areas of shape modeling research. Namely, this reluctance suggests that there exists \underline{a critical need} to develop hybrid systems that can provide 3D effects along with the convenience and expressive power of 2D. 

\begin{figure}[htbp!]
  \centering
  \begin{subfigure}[t]{0.33\textwidth}
\includegraphics[width=1.0\textwidth]{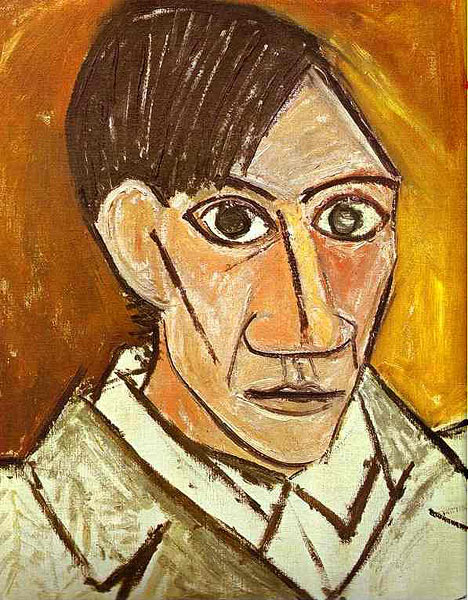}
  \caption{ An example of incoherent scenes: A cubist self-portrait by Pablo Picasso from 1907. }
  \label{fig_Picasso/original}
  \end{subfigure}
  \hfill
    \begin{subfigure}[t]{0.64\textwidth}
\includegraphics[width=0.530\textwidth]{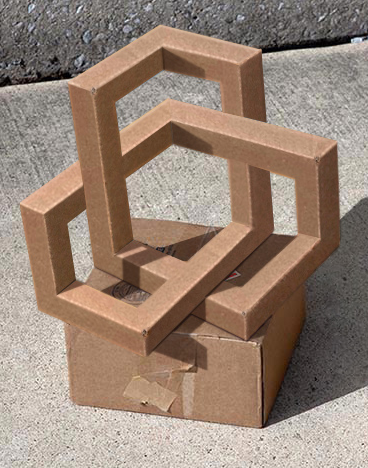}
\includegraphics[width=0.450\textwidth]{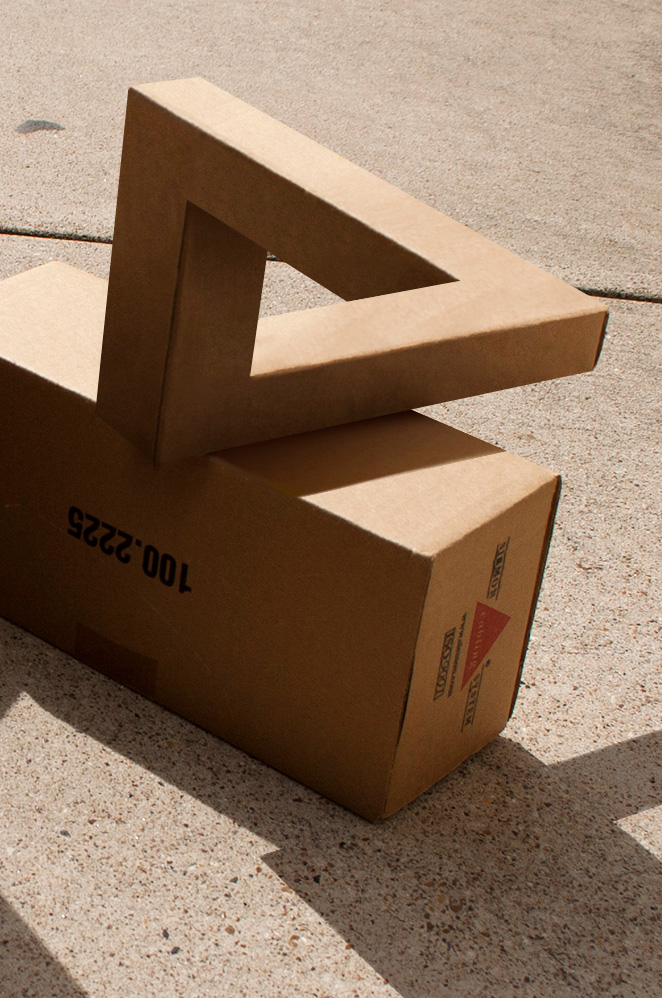}
  \caption{Examples of impossible objects: A hand-drawn composition of two impossible objects on photographs as composition class project.}
  \label{fig_impossible/34}
  \end{subfigure}
  \hfill
    \begin{subfigure}[t]{0.995\textwidth}
\includegraphics[width=0.560\textwidth]{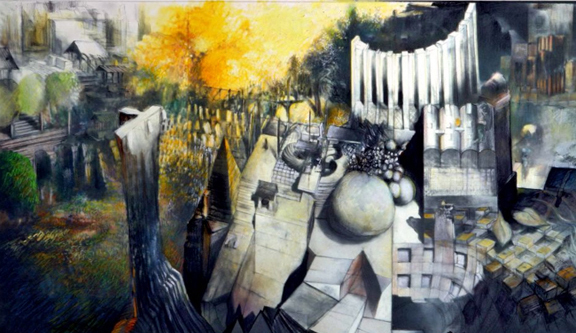}
\includegraphics[width=0.430\textwidth]{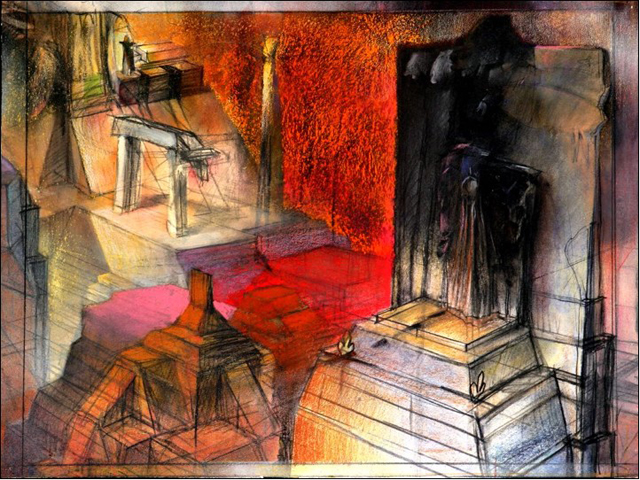}
  \caption{An example of inconsistent scenes: A landscape painting by Richard Davison from 2001. Davison intentionally introduced contradictory vanishing points in this painting.}
  \label{fig_impossible/davison1}
  \end{subfigure}
  \hfill
\caption{ Examples of static 2D pictorial documents that include incoherent, inconsistent, and impossible expressive depictions. For example, cubist artists create images based on their successive and subjective experiences in space and time \cite{gleizes1947,robbins1985} that result in incoherent structures. Our approach makes it possible to turn any of such structures into dynamic ones with re-renderable elements. }
\label{fig_teaser}
\end{figure}
  
In this paper, we present a framework for developing such {\em hybrid systems} that can support expressive depictions of impossible, inconsistent, and incoherent shapes and scenes. Although there exists a significant amount of research on non-(photo)realistic rendering (NPR), there has not yet been a comprehensive expressive depiction framework that can provide an integrated non-realistic approach for both modeling and rendering. There is, therefore, a need for a representation that is powerful enough to handle all types of expressive depictions from impossible renderings/shapes to incoherent or inconsistent renderings/shapes.

We envision a future in which static pictorial documents such as illustrations, paintings, and photographs are converted into dynamic re-renderable forms that can be accessible and continuously enriched by almost everybody. Our specific goal in this paper is to present an easy-to-use, easy-to-extend, and powerful framework that can provide a formal representation for such future applications. This new framework turns shape modeling into a 2D graphics application, and users can define shapes by painting images, creating illustrations, and photographing real objects.  

The key part of this framework is a mock-3D scene representation that consists of texture-mapped 2-complexes, and the key part of this representation is the textures that define non-conservative 2D vector fields along with thickness fields, which we call shape maps. Using shape maps, for any given mock-3D scene and a given 3D position, we can uniquely compute every 3D shape in the scene using rays emanating from the given position. These mock-3D scenes are view-dependent since the shapes of all objects in the scene depend on the positions of ray centers. Using these dynamically computed shapes, we can compute any illumination effect that requires geometry, such as shadows, reflection, and refraction, in real-time. 

\section{Mock-3D with Non-Conservative Vector Fields}

An important property of illustrations and paintings is that they rarely correspond directly to real 3D scenes; they are usually expressively depicted stylized representations and/or interpretations of real 3D scenes \cite{Gooch1998, Winkenbach1994, House2007}. Therefore, it is impossible to turn illustrations such as those shown in Figure~\ref{fig_teaser} into real 3D scenes, since the shapes in such pictures rarely correspond to real 3D shapes, the illumination is usually inconsistent, and the rendering is almost always expressive and stylistic. 
 
In paintings and illustrations, styles vary significantly from one artist to another. Like styles, shape, and illumination, inconsistencies are also introduced by artists ---usually on purpose--- since inconsistencies can make images interesting. And most importantly, if the fake 3D effects that 2D painting/editing provides are good enough, people will still tend to view these images as if they are in the 3D world. For example, the fact that the imperfect perspective in Figure~\ref{fig_teaser}(b) does not distract us from appreciating the picture. Richard Davison intentionally introduced contradictory vanishing points in this painting to demonstrate that humans do not consciously check for optical correctness \cite{davison2007}.

To bridge the gap between 2D painting and 3D rendering, we need mock-3D shapes that are not 3D but appear 3D. In current practice, there has been some use of mock-3D representations in the form of normal and depth maps \cite{wang2014global,gonen2016quad,wang2014qualitative,akleman2017cos,akleman2022dynamic,akleman2023webbased}. However, these representations, such as normal maps, usually do not correspond to thoroughly impossible shapes. Depth maps are essentially Bas-Reliefs \cite{Weyrich2007}. They cannot represent impossible shapes since their gradient can only produce conservative vector fields. Normal maps are usually produced as conservative vector fields, which are constructed as gradients of height functions \cite{wiki}. In both cases, there is always a unique constructible geometry \cite{Fattal2002}. 

 \begin{figure}[ht]
 \begin{tabular}{cccccc}
 \includegraphics[width=0.320\textwidth]{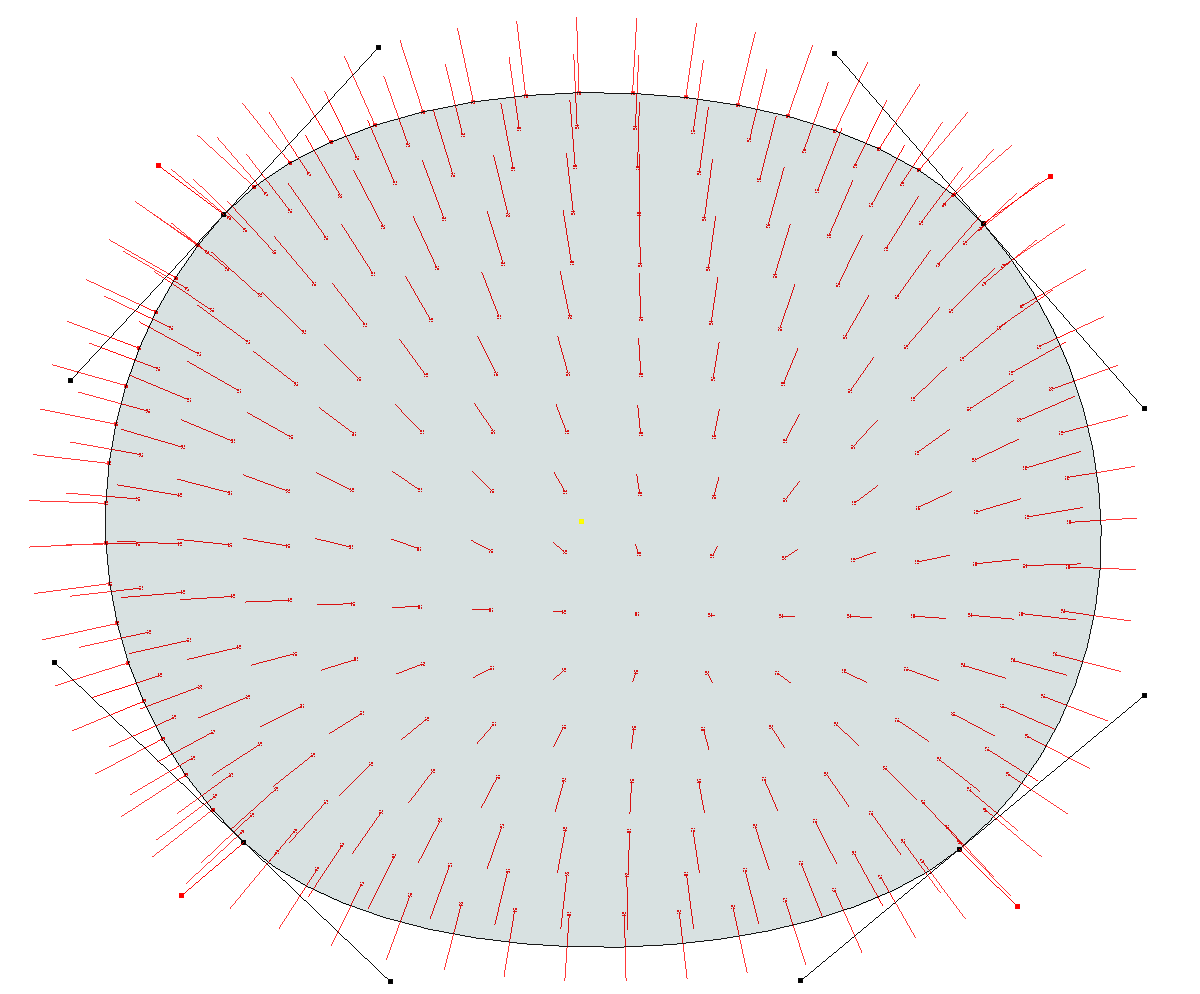}&
 \includegraphics[width=0.320\textwidth]{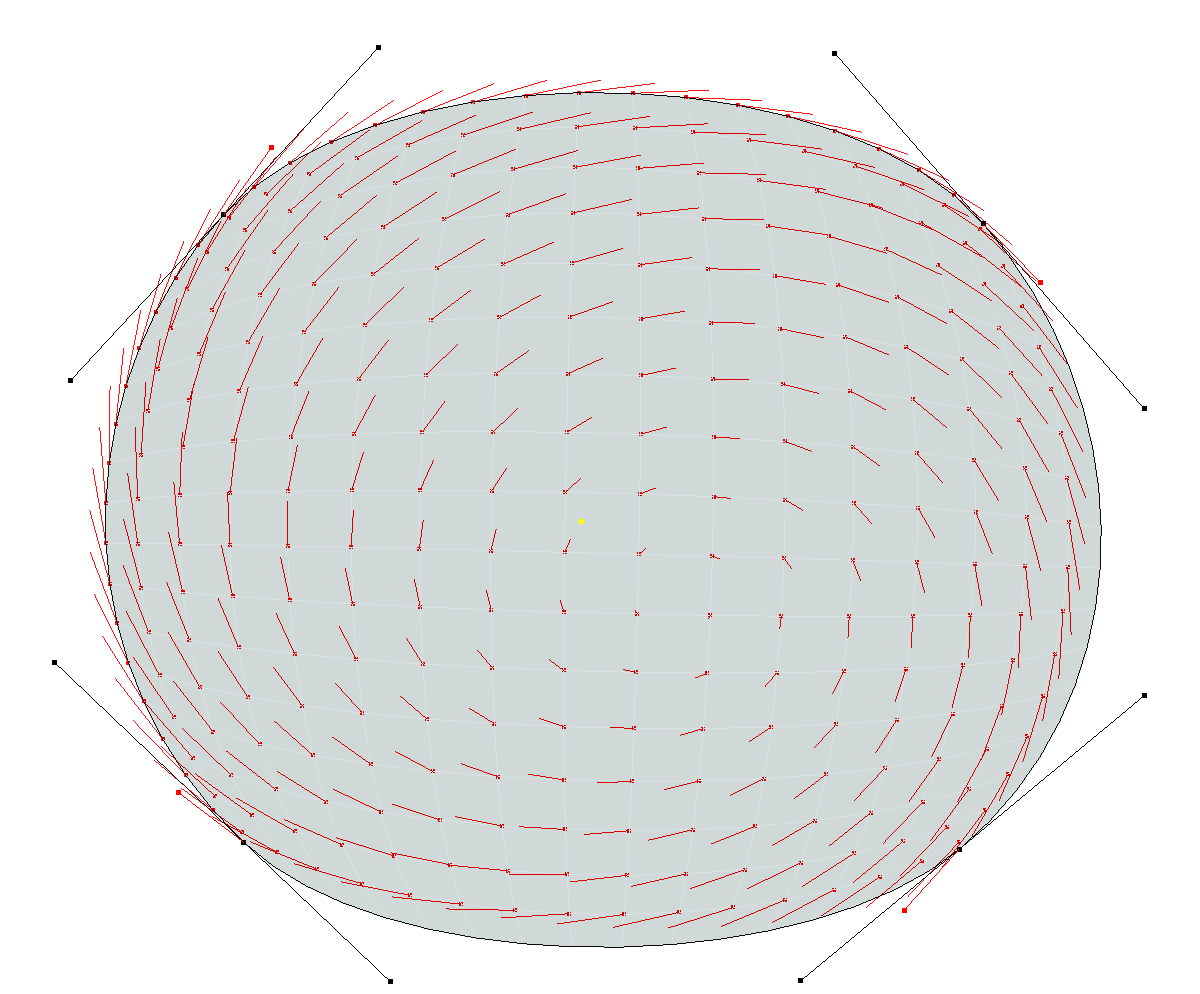}&
 \includegraphics[width=0.320\textwidth]{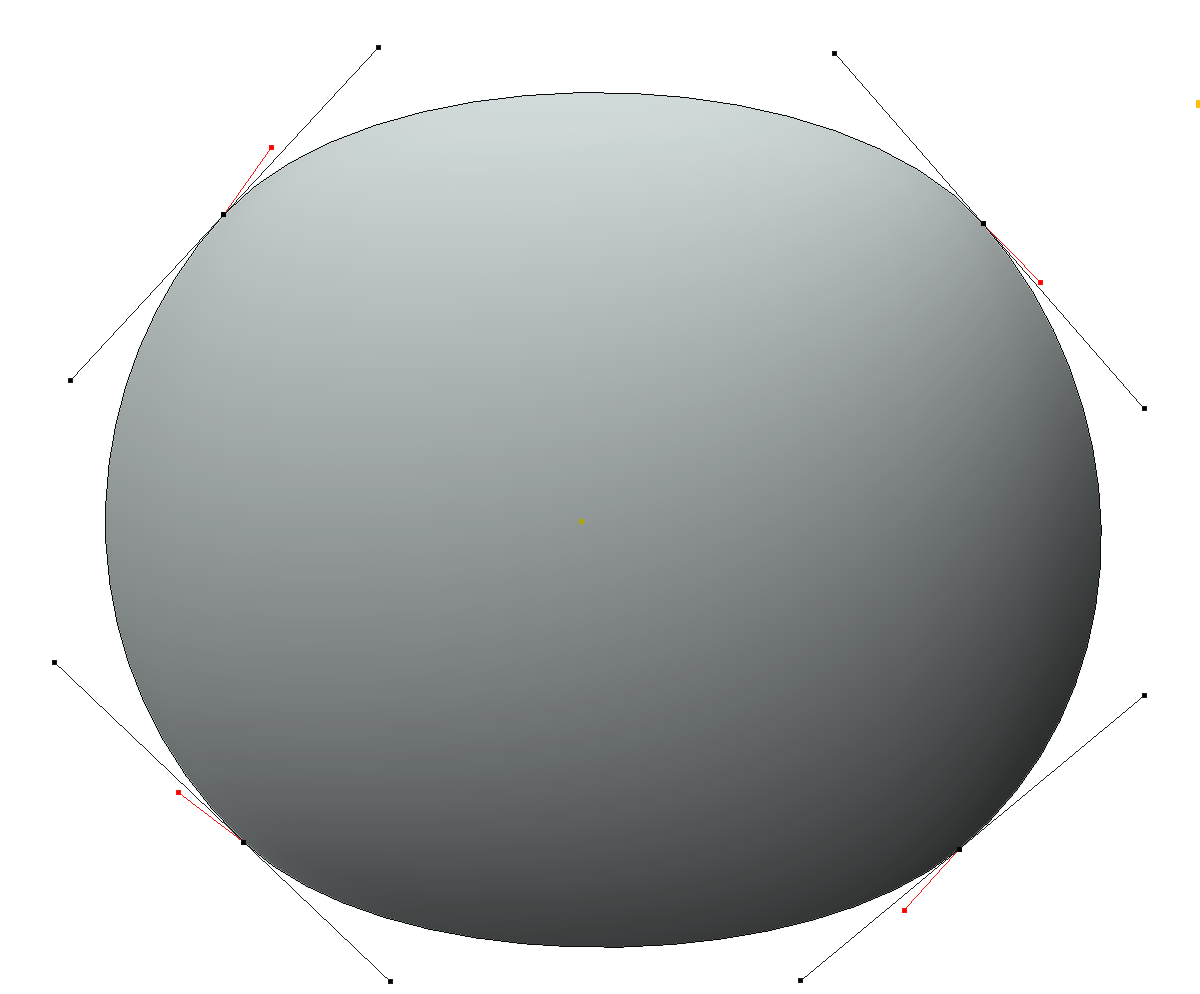}\\
  (a) Conservative field&
   (b)  Non-conservative field &
 (c) B\&W Rendering of (b)  \\
 \end{tabular}
 \caption{   Examples of (a) conservative and (b) non-conservative fields. (c) demonstrate that non-conservative vector fields can produce realistic-looking rendering by converting non-conservative fields to view-dependent geometry. More interestingly, such non-conservative fields can rotate the objects behind them through refraction. }
 \label{fig_vectorfield2}
 \end{figure}

In this paper, we propose using 2D vector fields to construct representations that truly mock 3D geometry. The advantage of random 2D-vector fields is that they do not necessarily come from gradients of height fields. Therefore, they are not necessarily conservative. If the field is non-conservative, there is no corresponding height field, and, as a result, we have a mathematical representation that does not correspond to any real shape. Non-conservative vector fields are used to conceptualize impossibility in shapes such as the never-ending staircase in Escher's "Ascending and Descending" \cite{wiki2}. In other words, the words "impossibility", "inconsistency" or "incoherency" really refer to global consistency that can be introduced by a nonconservative vector field. 

Fortunately, this global inconsistency does not prevent us from locally reconstructing height fields. In fact, for any given line in 2D, we can always construct a slice of a height field from any given vector field using a simple line integral.  If we choose a set of rays emanating from the same point, we can then construct the whole height field in 2D. The reconstructed height field, of course, depends on the point from which the rays emanated. These shapes are therefore view-dependent, which is, in fact, also a desired property in cartoon animation \cite{Rademacher1999}. 

The problem is that 2D-vector fields in the plane can provide only mock height fields. Even if we map them on 3D surfaces, we can only obtain mock displacement fields. Neither of them thoroughly provides boundaries for 3D solids. We, therefore, need to give them a volume to turn them into mock-3D solid objects. This can be done using two vector fields, one for positive displacement and another one for negative displacement. A simple solution is simply to add a thickness field to create the second displacement. In other words, both two-sided mock-3D-displacements can be described by only three numbers, which can then be provided by single three-color --RGB-- images. We use the term {\em``shape map''} to describe the images that provide this two-sided mock-3D displacement information.  

\begin{figure}
\vspace{-0.25in} 
\centering
\subfloat[A Mock-3D scene.]
{\includegraphics[width=0.20\textwidth]{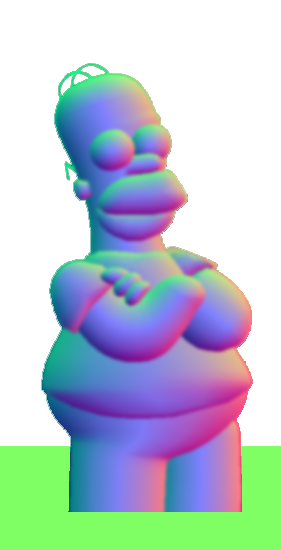}}\hfill
\subfloat[  Shadow from front light.]
{\includegraphics[width=0.39\textwidth]{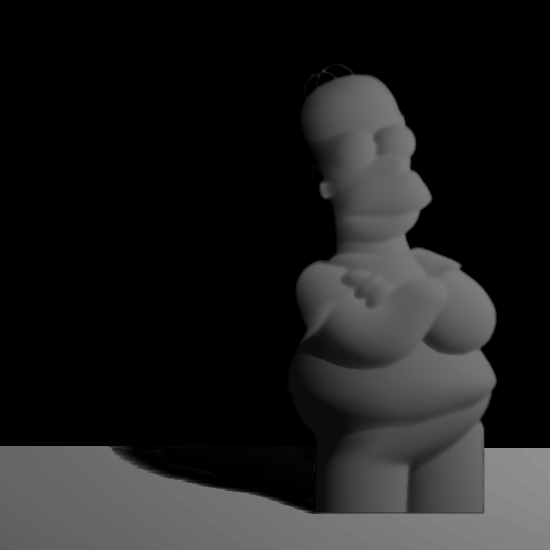}}\hfill
\subfloat[ Shadow from  side light. ]
{\includegraphics[width=0.39\textwidth]{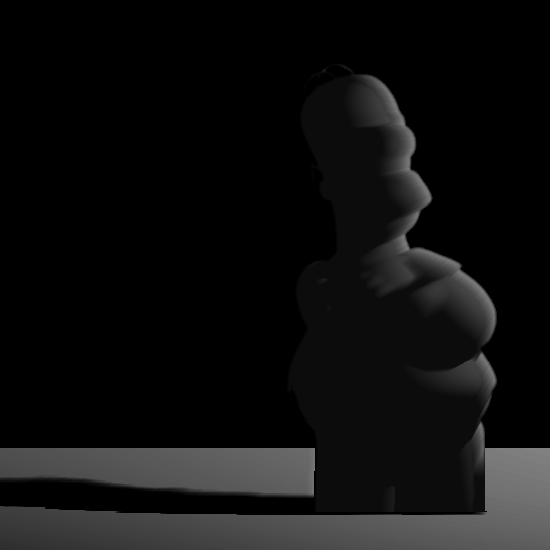}}
\caption{   An example of local and global shadows cast by a shape map mapped on a planar billboard. The mock-3D scene consists of two texture-mapped planar rectangles. Note that global shadow is volumetric even when light is in the same plane as the billboard. }
\label{fig_homer}
\vspace{-0.125in}
\end{figure} 

Figure~\ref{fig_homer} shows a mock 3D scene that consists of only two texture-mapped planar rectangles. The colors in ~\ref{fig_homer}(a) provide two-sided mock-3D-displacement information.  Using this information, we can obtain global shadows as if the planar shapes had solid volumes. However, the straight line created by the intersection of the two planes creates a visual distraction. To avoid this problem, we need more flexible structures than planar quadrilaterals. 

\section{Mock-3D Scenes with 2-Complexes}

\begin{figure}[htp]
\centering
\subfloat[A single 2-manifold mesh with boundary.]
{\includegraphics[width=0.26\textwidth]{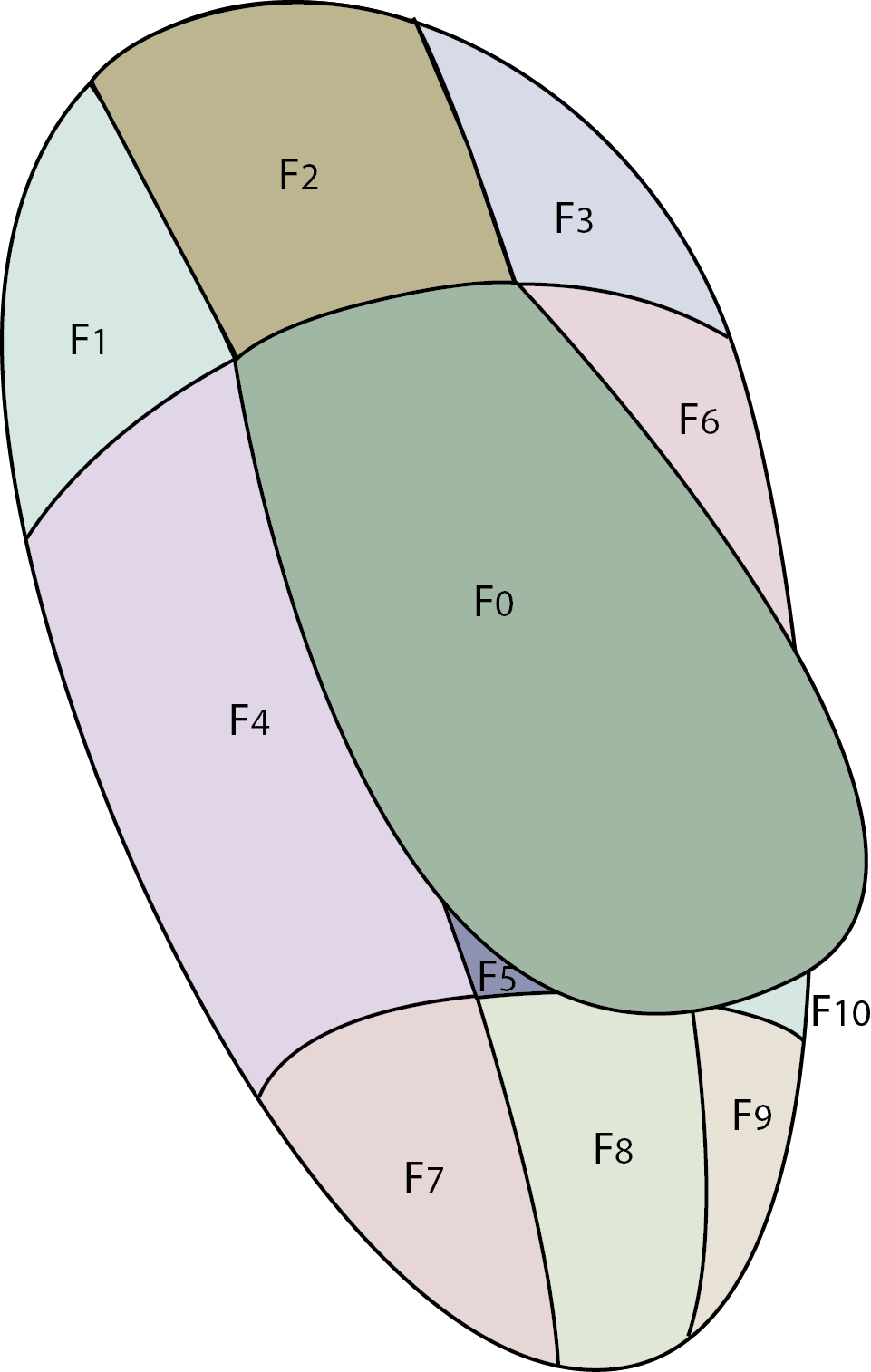}}\hfill
\subfloat[ A group of two 2-manifold meshes with boundaries.]
{\includegraphics[width=0.42\textwidth]{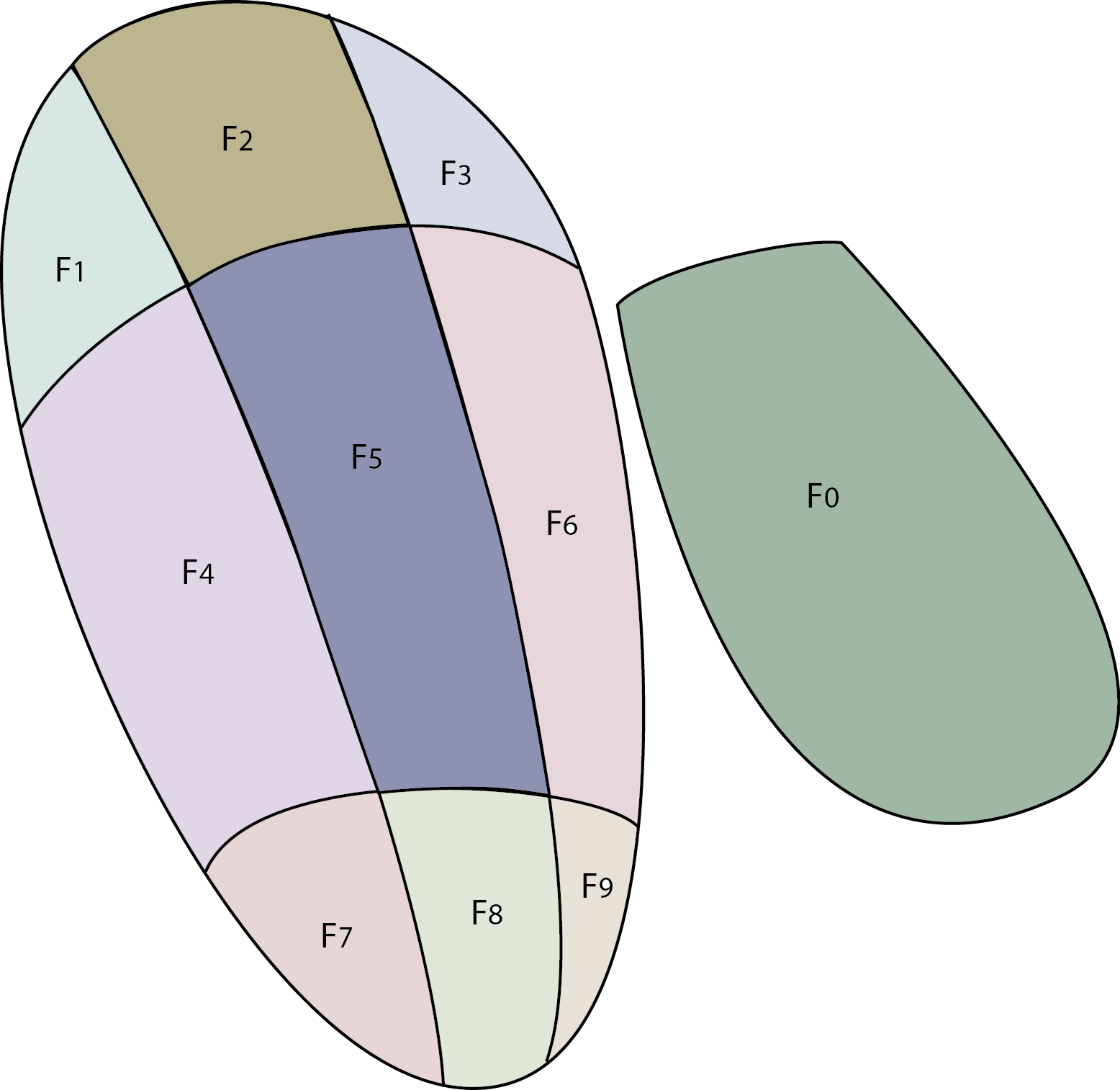}}\hfill
\subfloat[ A single 2-complex mesh.]
{\includegraphics[width=0.26\textwidth]{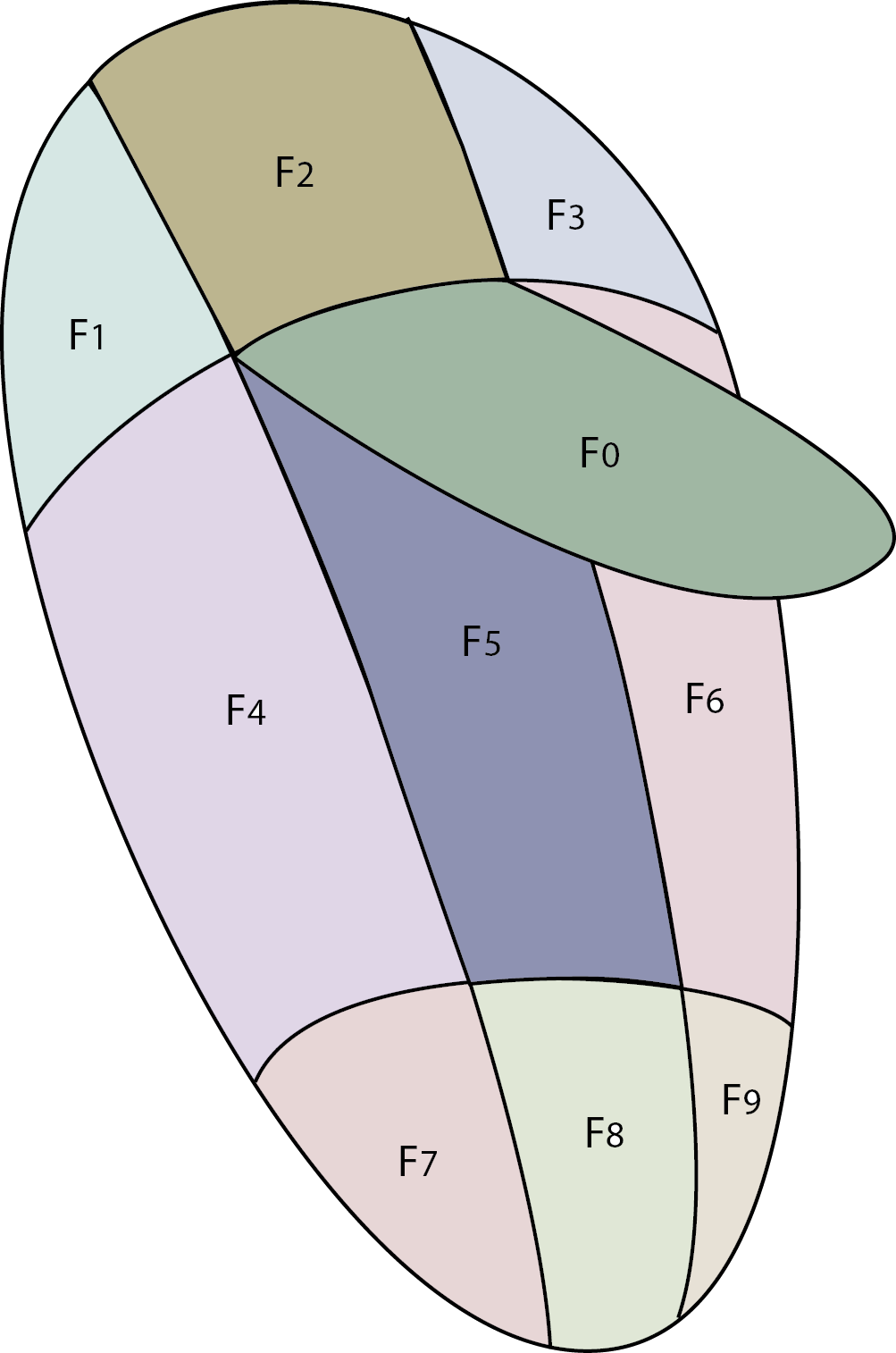}}
\caption{  An example that demonstrates the advantage of 2-complexes in representing single objects. (a) shows a cartoon face represented by a single 2-manifold mesh with a boundary. This requires a complex mesh structure that should change with shape. For example, in (a) $F_4$ and $F_8$ are pentagonal patches; and $F_5$, $F_6$ and $F_{10}$ are triangles and the rest of the patches are quads. If we make the nose slightly smaller, we need to reconstruct the mesh. Using layers, it is possible to obtain general structures that do not require restructuring and/or remeshing. For example, in (b), regardless of the size of the nose, we can keep every patch as a quad. In this case, the problem is that we lose the information that the faces $F_0$ and $F_2$ share an edge. On the other hand, 2-complexes provide the best of both worlds and can describe that $F_0$, $F_2$, and $F_5$ share the same edge.}
\label{fig_2complex}
\end{figure}

To utilize this approach in a general setting, we propose to construct mock-3D scenes with 2-complexes, which can be represented using the shape algebra recently introduced \cite{Akleman2015mod1}. 2-complexes in 3D can include deformed planes, curves, and their connections (see Figure \ref{fig_2complex}). Z-depth deformations introduced by Gershon Elber to represent impossible objects are particularly useful in this case, since they can deform objects without changing the visible parts \cite{Elber2011}.  Deformations in the Z depth can also effectively provide local layering \cite{McCann2009}, which is important to include cases that cannot be handled by simple ordering, such as knots, links, and handshakes. We will refer to each 2-complex as a ``layer'' just to be consistent with the standard terminology used in image manipulation, as they will appear as layers to the most casual users.

Using 2-complexes to represent single 2D objects may seem to be an overkill, and one may think that every 2D object can be represented simply by using a 2-manifold mesh with boundary. Figure~\ref{fig_2complex}(a) shows how impractical it can be to use a single 2-manifold mesh with a boundary to represent a 2D object. As shown in the figure, if we use such a structure, we need to change the mesh structure of the 2-manifold with any change in any part of the object. The most common solution in Computer Graphics for such problems is to use groups.  For example, if we want to represent a human, we can have layers of each part of the body, such as arms, legs, body, and head. For each part, we can also have subparts, which are also defined as sublayers. Then, the whole structure is organized into groups.

One problem in using groups: we lose connectivity information available in using a single 2-manifold mesh. If we want to keep connectivity information, we need 2-complexes. For instance, if more than two polygons share an edge, we cannot use 2-manifolds, by definition. On the other hand, two complexes can easily provide the information that the three faces, $F_0$, $F_2$, and $F_5$, share the same edge.  Until recently, it was hard to use 2-complexes since there was no strong data structure to represent non-simplicial 2-complexes. Fortunately, a recently introduced general framework for representing 3-manifold decompositions can also be used to represent 2-complexes with some minor modifications \cite{Akleman2015mod1}. 

One significant advantage of 2-complexes is to develop templates for common objects such as humans, animals, chairs, or planes. These 2-complex templates can be much simpler than 3D shapes and, therefore, easy to use for most people. These 2-complexes, i.e. layers, are dynamic objects that can help us re-render the image. Therefore, each layer should consist of several components, each of which can be considered as textures (raster or vector) that are projected to the layers. We call these components ``channels'' consistent, again, with the standard terminology used in image manipulation. On the other hand, the term channel in this case will refer to entities that are more general than simple color channels. For example, one channel should be a shape map that provides shape information. These shape maps that contain 2D vector field and thickness information, as mentioned earlier, help us to turn these 2-complexes into mock-3D shapes by providing two-sided mock-3D-displacements. 

\begin{figure}
\vspace{-0.25in} 
\centering
\subfloat[Original. ]
{\includegraphics[width=0.32\textwidth]{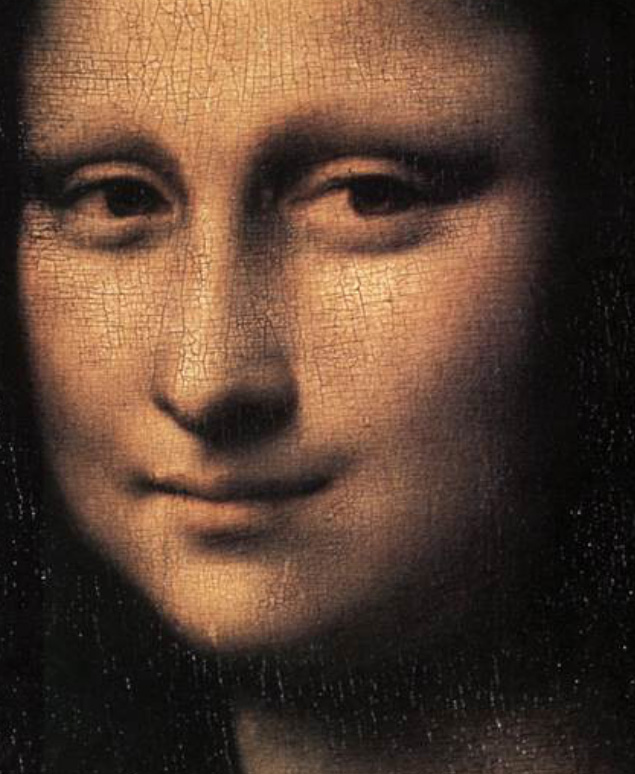}}\hfill
\subfloat[Relighting. ]
{\includegraphics[width=0.32\textwidth]{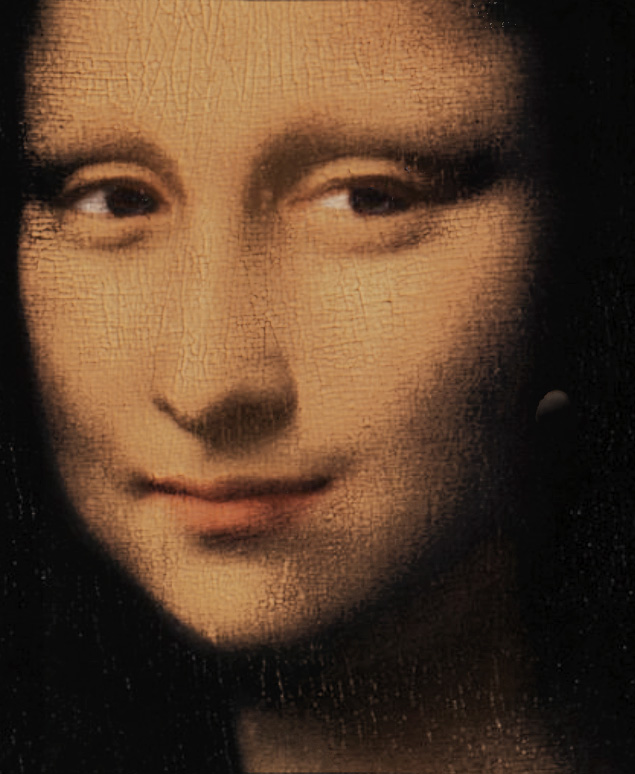}}\hfill
\subfloat[Relighting. ]
{\includegraphics[width=0.32\textwidth]{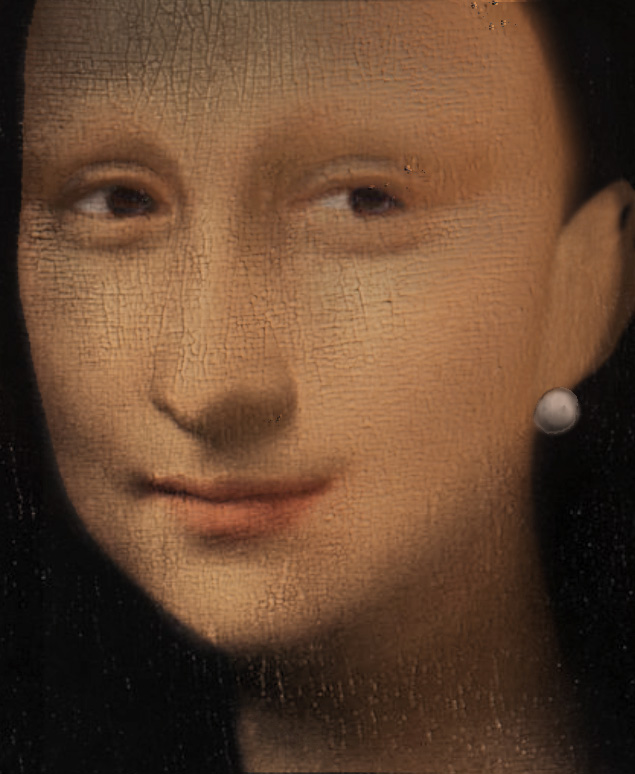}}
\caption{  Relighting Mona Lisa: an example of face relighting through channel extraction by removing shadows and shading. Note that (c) and (d) show an artificially added pearl earring as an homage to Vermeer's painting ``The Girl with Pearl Earring''. }
\label{fig_MonaLisa}
\end{figure}

Using channels, we can also provide material properties to control how the final image should be rendered. For example, Figure~\ref{fig_homer} does not have any material information and the images in Figure~\ref{fig_homer}(a) and (b) are simply diffuse illumination. Figure~\ref{fig_MonaLisa} demonstrates how material information can be used to obtain a particular style.

\begin{figure}[ht]
\begin{center}
\begin{tabular}{ccccccc}
 \includegraphics[width=0.32\textwidth]{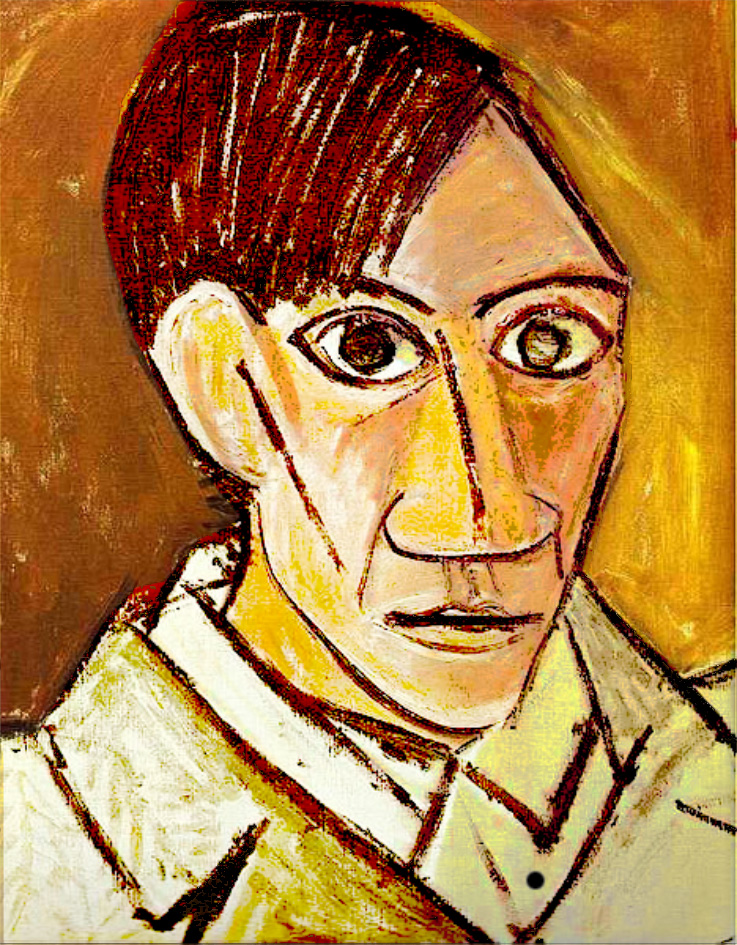}&
 \includegraphics[width=0.32\textwidth]{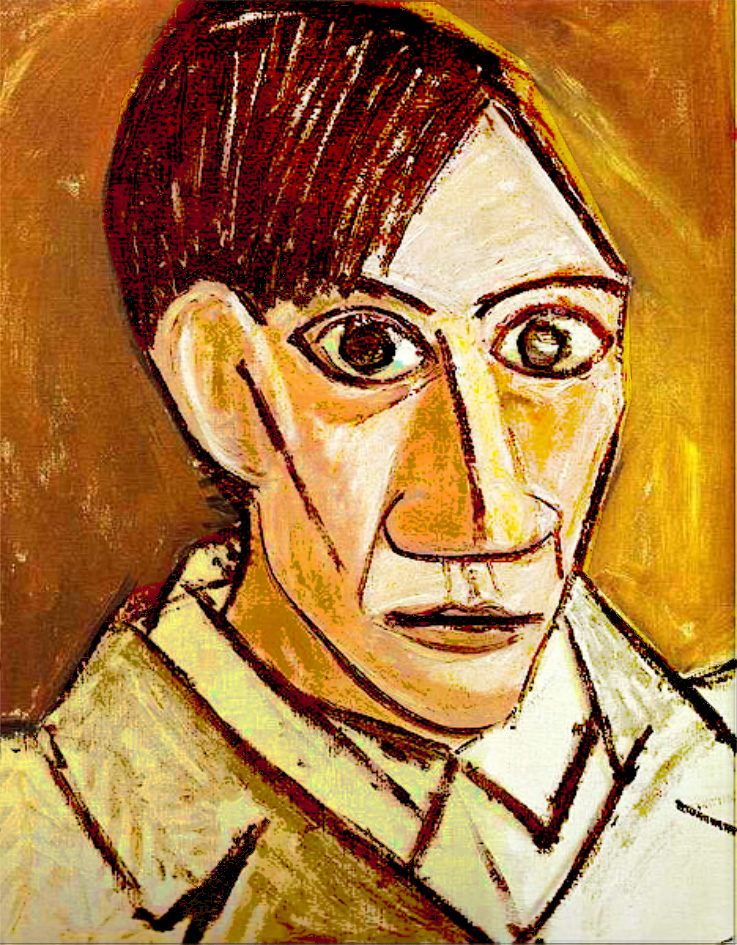}&
 \includegraphics[width=0.32\textwidth]{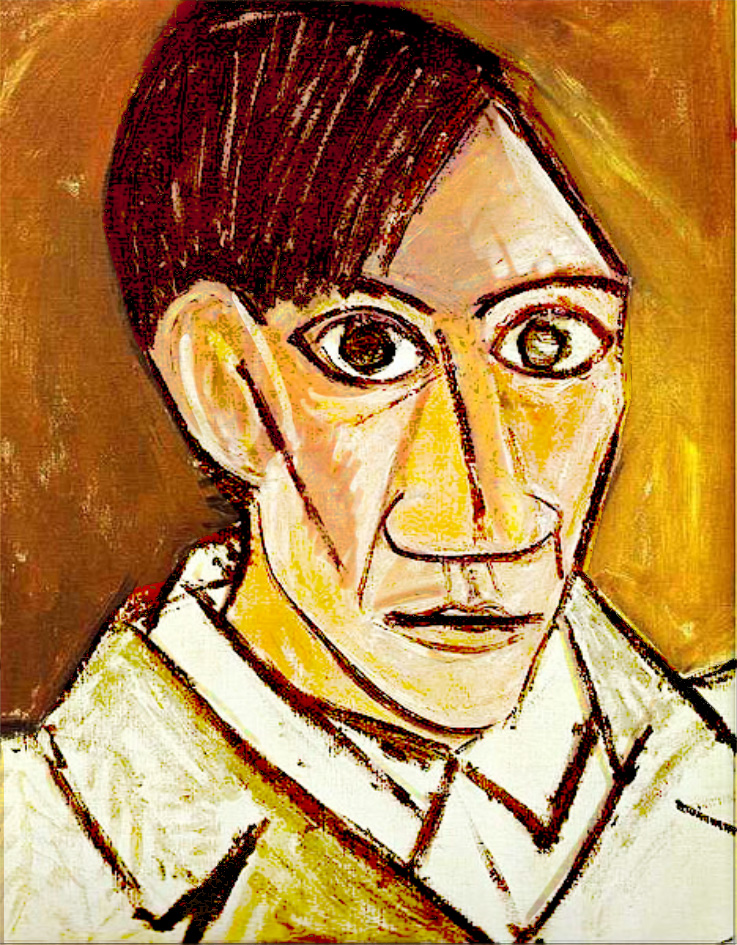}\\
\end{tabular}
\end{center}
\caption{ Relighting Picasso's self-portrait. Note that although this image is intentionally flattened and does not correspond to any real shape, it is still possible to illuminate it with our approach.}
\label{fig_Picasso}
\end{figure}

\section{Related Work}

There currently exist two representations that are related to ours: bas-reliefs and normal maps. However, both of them correspond to real shapes that can exist in 3D. 
In this work, we present a fuzzy and view-dependent representation that is suitable for global illumination while providing all the representational powers of both bas-reliefs and normal maps.

Bas-reliefs are sculptures that can be viewed from many angles with no perspective distortion as if they are just images. In other words, the perspective transformation is embedded in bas-relief sculptures \cite{Weyrich2007}. One problem with bas-reliefs for 2D artists is that their construction is still a sculpting process. This may not be suitable for illustrators and painters who are not interested in sculpting shapes.

\begin{figure}[htbp]
\begin{center}
\begin{tabular}{cccccc}
 \includegraphics[height=0.22\textheight]{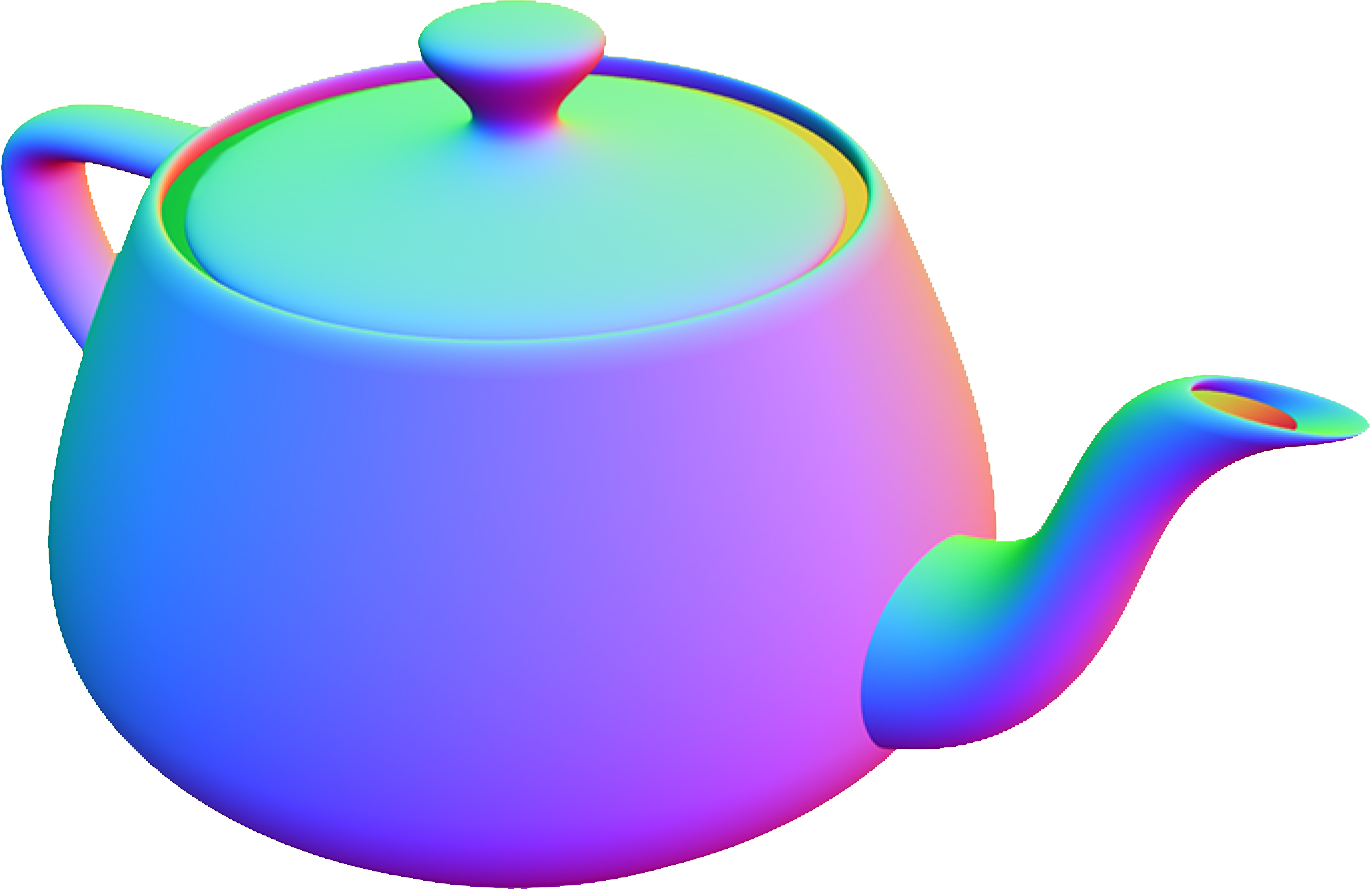}&
 \includegraphics[height=0.22\textheight]{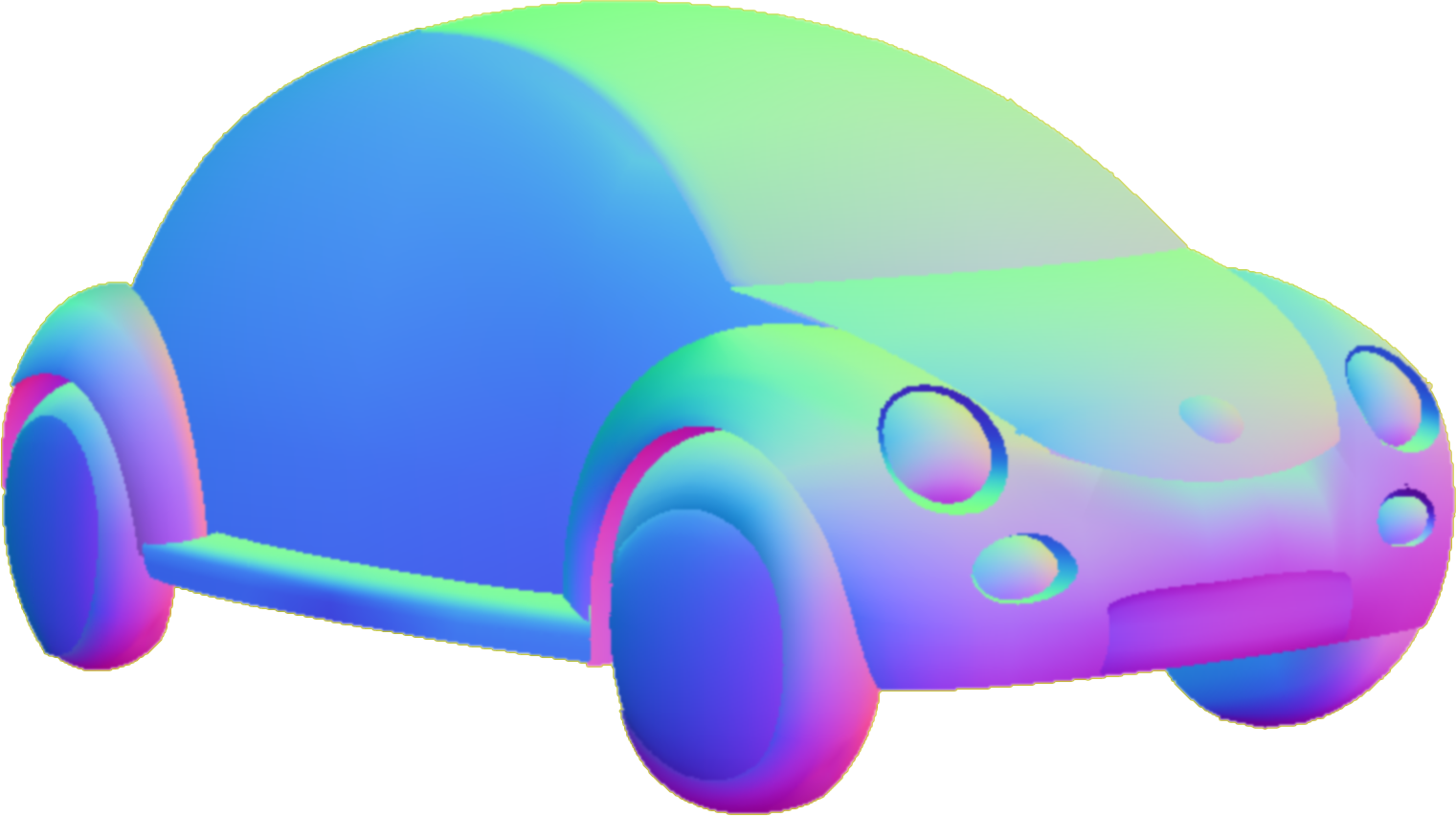}\\
Lumo & CrossShade
\end{tabular}
\end{center}
\caption{ Examples of normal maps generated using sketch-based modeling programs. Normal maps are designed to correspond to conservative vector fields. Therefore, there always exists a unique bas-relief corresponding to a normal map.}
\label{fig_Normal}
\end{figure}

Normal maps became popular as soon as they were introduced in 1998 \cite{Cohen1998}. Although they are mainly used as texture maps to include details in polygonal meshes, they can be used directly as shape representations by embedding perspective information, as shown by Johnston \cite{Johnston2002}. He developed a sketch-based system, called Lumo, to model normal maps by diffusing 2D normals in a line drawing. Since then, only a few groups have investigated the potential use of normal maps as a shape representation, such as \cite{Okabe2006, Bezerra2005, Winnemoeller2009, Shao2012}. Sun et al. \cite{Sun07} introduced Gradient Mesh to semi-automatically and quickly interpolate normals from edges, and Orzan et al. \cite{Orzan09} calculate a diffusion from edges by solving the Poisson equation. Sykora et al. \cite{Daniel09} proposed the Lazy-Brush, which can propagate scribbles to accelerate the definition of constant-color regions. Finch et al. \cite{Finch11} build thin-plate splines that provide smoothness everywhere, except at user-specified tears and creases. The underlying splines are used to interpolate the normals. 

Wu et al. \cite{Wu07} proposed a shape palette, where the user can draw a simple 2D primitive in the 2D view and then specify its 3D orientation by drawing a corresponding primitive. This method also performs diffusion using a thin-plate spline. Recently, Shado et al. \cite{Shao2012} developed CrossShade, another sketch-based interface to design complicated shapes such as normal maps. They used an explicit mathematical formulation of the relationships between cross-sectional curves and geometry. The specified cross-section is used as an extra control point to control the normals. Vergne et al. \cite{Vergne2012} introduce surface flow from smooth differential analysis, which can be used to measure smooth luminance variations. Therefore, the author also proposes drawing shadows and other shading effects. 

The issue from our perspective is that normal maps are designed to correspond to conservative vector fields, and there is, therefore, always a unique bas-relief corresponding to a normal map. For example, Sykora et al. \cite{Sykora2014} developed a user-assisted method to convert normal maps into Bass-Reliefs that can provide correct shadows in a commercial renderer, but this approach will fail if the normal maps do not correspond to Bas-Reliefs that can have explicitly meaningful 3D geometry.

\section{Shape Maps: Images as Mock-3D Shapes}

Let $F_0(x,y,\theta)$ and $F_1(x,y,\theta)$ denote two functions of $ \mathcal{D} \times [0, 2 \pi]$ to $\Re$ with
$(x,y) \in \mathcal{D}$ and $\theta \in [0, 2 \pi]$. Then, a view-dependent and 2-sided mock 3-D shape can be defined using the following inequality: $$ S = F_0 (x, y,\theta) \geq z \geq  F_1(x,y,\theta) $$
where $z \in \Re$.

The first difference with traditional bas-reliefs is that these Mock-3D shapes are two-sided. In this definition, the two-sidedness of the shape directly comes from the usage of two functions $F_0$ and $F_1$. The second and more important difference from bas-reliefs is that the shape is also a function of angle $\theta$. In other words, this definition allows us to have a view-dependent geometry since the shape depends on the angle $\theta$. View dependency is a desired feature in artistic applications \cite{Rademacher1999}. It also provides a fuzzy definition for shapes that can be particularly useful for representing impossible shapes. The only problem with this general structure is that it is hard for 2D artists to create these two essentially 3D functions. We therefore present shape maps, which are represented as images. We demonstrate that shape maps can be used to construct these two functions.

Shape map images encode two types of information: a 2D vector field and a thickness map. Let $\mathcal{D} \in \Re^2$ denote the 2D vector field of the domain to be defined, and let $p=(x,y) \in \mathcal{D}$ denote two coordinates of the map. Let $(N_0(p),N_1(p))$ denote the 2D vector field $N_0 : \mathcal{D} \rightarrow [-1,1], $ and $N_1 :\mathcal{D} \rightarrow [-1,1]$.  Thickness maps are also defined over the same domain of the 2D vector field where $T(p)$ denotes a thickness map as $T: M \rightarrow [0,1]$. 
One significant advantage of having only three variables for shape maps is that we can readily convert them into Low Dynamic Range (LDR) images and save them using any common image format, which can easily passed to GPU like a normal map. Let an LDR image on $M$ be denoted by $c(x,y) = (r(x,y),g(x,y),b(x,y))$ where $c: \mathcal{D} \rightarrow [0,1]^3$. The conversion from $(N_0,N_1)$ and $T$ to $(r,g,b)$ is given as $(r=0.5(N_0+1),g=0.5(N_1+1),b=T)$. We use the opacity channel $\alpha$ to describe the domain of the function $\mathcal{D}$. If $(x,y) \notin \mathcal{D}$, we choose $\alpha=0$. Of course, aliasing needs to be avoided in this case by appropriately sampling each pixel.

 \begin{figure}[ht]
 \begin{tabular}{cccccc}
 \includegraphics[height=0.32\textheight]{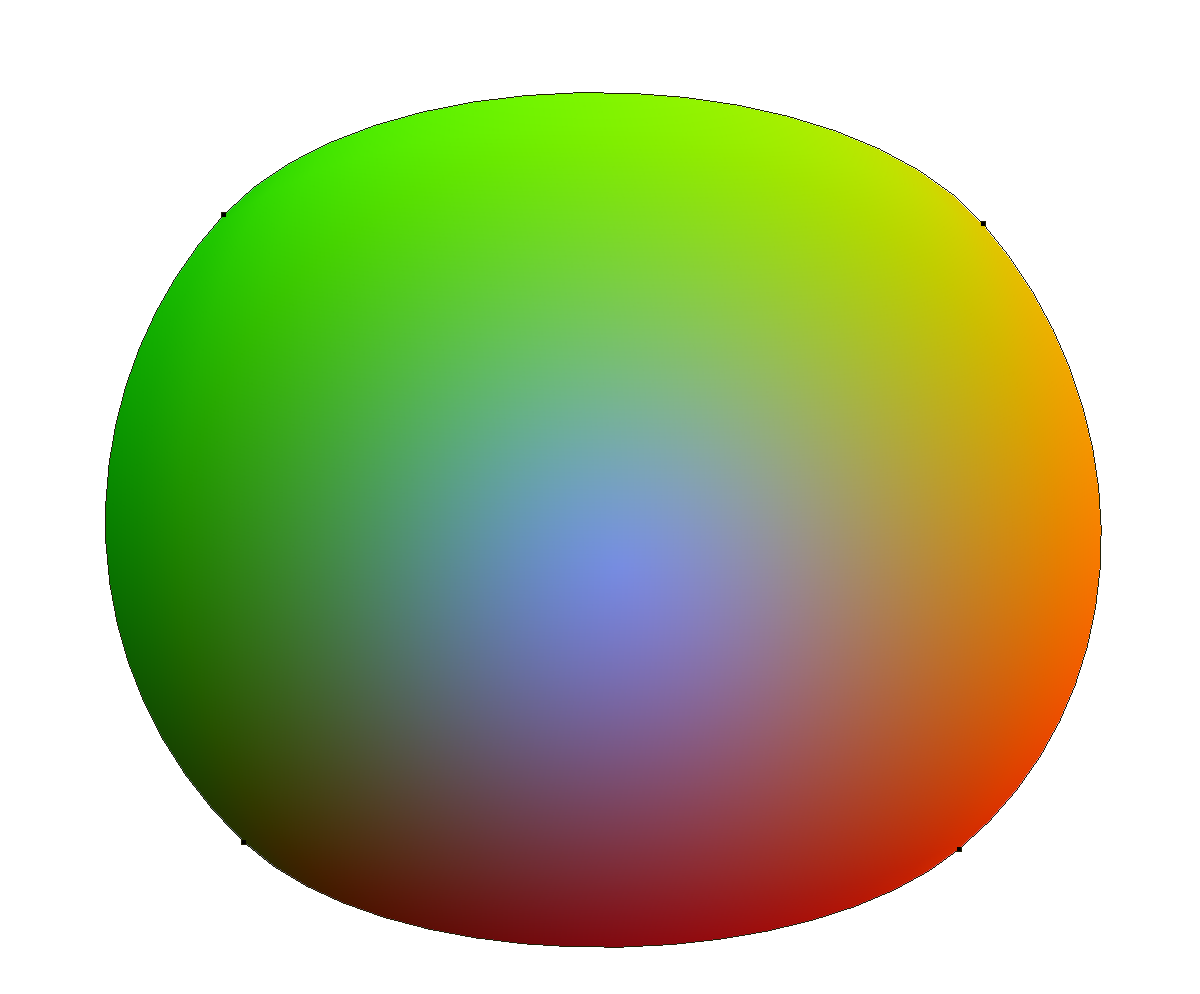}&
 \includegraphics[height=0.32\textheight]{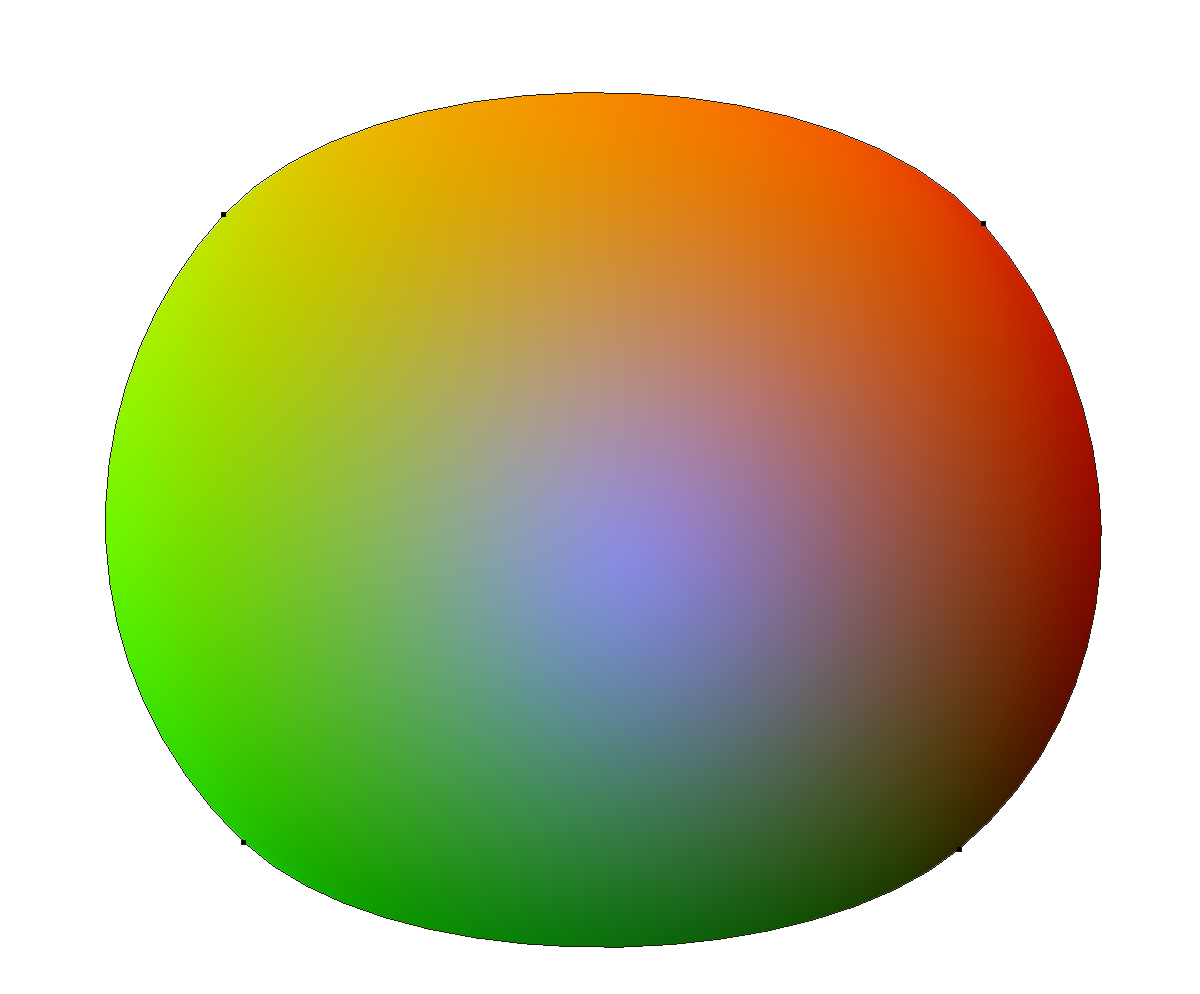}\\
  (a) Shape map of a conservative field&
   (b) Shape map of a non-conservative field  \\
 \end{tabular}
 \caption{   Examples of shape maps of conservative and non-conservative fields from Figure~\ref{fig_vectorfield2}. In these examples, the thickness is zero in the boundaries and the largest in the center. Note that they do not look too different visually. }
 \label{fig_vectorfield3}
 \end{figure}

Using these maps, the functions $F_0$ and $F_1$ are computed as a summation of the two line integrals of 2D gradient fields that are obtained directly from 2D vector fields and displacement maps as
$$F_0(p_1,\theta) =s_0 G_0(p_1,\theta) + s_1  G_1(p_1,\theta) \label{eq:gsum}$$
$$F_1(p_1,\theta) = F_0(p_1,\theta) - s_2 T(p) $$
where
$$G_0(p_1,\theta) =  \int_{p_0}^{p_1} N_0 (p(t)) \cos \theta dt + N_1 (p(t)) \sin \theta dt$$
$$G_1(p_1,\theta) = \int_{p_0}^{p_1} \frac{1}{n}  \lfloor n \frac{\delta Z(p(t))}{\delta t} + 0.5\rfloor  dt \label{eq:integral}$$
with $p_0=(x_0, y_0)$ is the starting point of the integral, which is computed as the intersection of the ray starting from  $p_1=(x_1, y_1)$ in the direction of $(-\cos \theta, -\sin \theta t)$ with the boundary of the domain $\mathcal{D}$ and $p(t)=p_0 + (\cos \theta, \sin \theta) t $ and $\lfloor x \rfloor$ is the floor function that return largest integer smaller than $x$, $n$ is an integer, quantization term and $s_0, s_1, s_2 \in [0,1]$ are scale parameters. 

If the 2D vector field $(N_0,N_1)$ is conservative, then $G_0(p_1,\theta)$ is independent of $\theta$. In other words, this integral provides all continuous function bas-reliefs when the 2D vector field is conservative. If the 2D vector field is not conservative, then the integral is dependent on $\theta$, but it is still uniquely defined. The resulting function is continuous in the direction of $\theta$ and may not necessarily be continuous in other directions. Therefore, it turns a 2D vector field defining an impossible shape into a fuzzy geometry, and a thickness map provides the back side of the shape. Shape maps, which are mapped onto layers of 2-complexes, can be considered ``extended billboards". 

These texture-mapped layers can easily be used to create scenes that allow for local and global illumination effects. We first project all rays and shapes to a 2D plane and we compute geometry in fragment shader dynamically.   The dynamically computed geometry is used to compute global illumination effects, such as ambient occlusion, shadows, and refraction. On the other hand, for effects that only require surface normals, such as diffuse and specular reflection, dynamic computing surface normals from the local structure of geometry can be an overkill.  For a simple alternative, it is possible to rebuild normal vectors using the unit vector property as $(sN_0, sN_1, \sqrt{1-s^2 N_0^2-s^2 N_1^2})$ where $s \in (0,1]$ is another user-defined control that is used to scale the values of $N_0$ and $N_1$ to $sN_0$ and $sN_1$. This operation, in effect, changes the effective depth of a corresponding bas-relief if there exists such a height field. In other words, this value indicates how flat we want to make the corresponding bas-relief if it exists. Moreover, choosing $s\leq 1/\sqrt{2}$ always guarantees $1-s^2 N_0^2-s^2 N_1^2<0$.

\section{Creating Shape Maps}

Shape maps have some visual and conceptual similarities to normal maps. This similarity is useful since we can directly use normal maps as shapes if necessary. Moreover, the blue component of normal maps can still provide an acceptable thickness map. Despite this similarity, shape map images are usually more colorful than normal maps since (1) we allow non-conservative fields and (2) we use blue color for thickness information. In practice, any image can be used as a shape map. The main advantage of allowing any image as a shape map is that artists can create these maps directly by painting or rendering an image, by taking a photograph, or through illustration. 

\section{Painting or Rendering Shape Maps \label{section:RSM}}

To paint shape descriptors, artists can imagine an object that is illuminated with parallel red light from the left side and with parallel green light from the top. By ignoring shadows, they paint an image based on how much red and green light they want to see in every pixel. For instance, a pixel color red = 0.95 and green = 0.75 means that the artist wants 95\% of the light from the left and 75\% of the light from the top to illuminate that particular pixel \footnote{This information can be interpreted as local scattering information from left and right lights This could be one of the additional reasons why non-conservative fields can produce realistic looking renderings by implicitly providing subsurface scattering from sides.}.  Moreover, unlike normal maps, it cannot be guaranteed that the sum of their squares will be smaller than $1$ as in this example. This is not a problem for estimating surface normals or overall shape as previously discussed. 

\begin{figure}[htbp]
\begin{center}
\begin{tabular}{cccccc}
 \includegraphics[height=0.21\textheight]{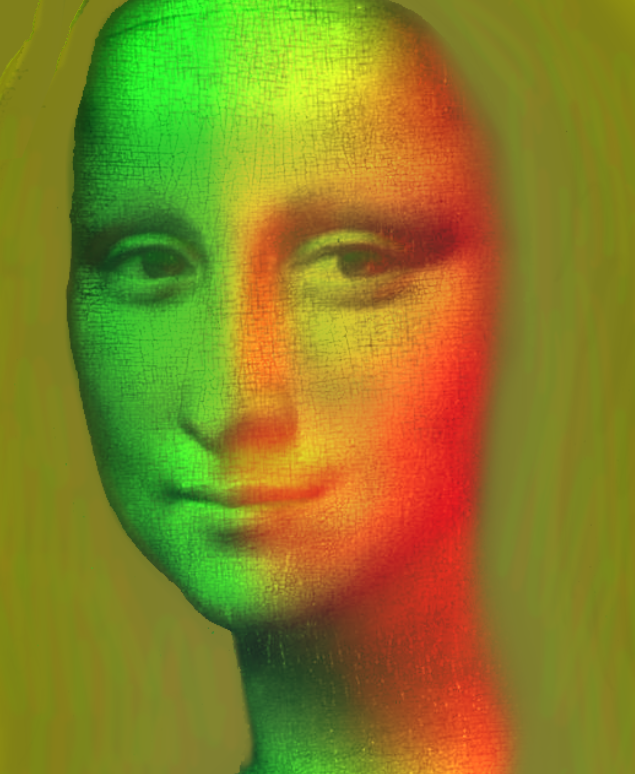}&
 \includegraphics[height=0.21\textheight]{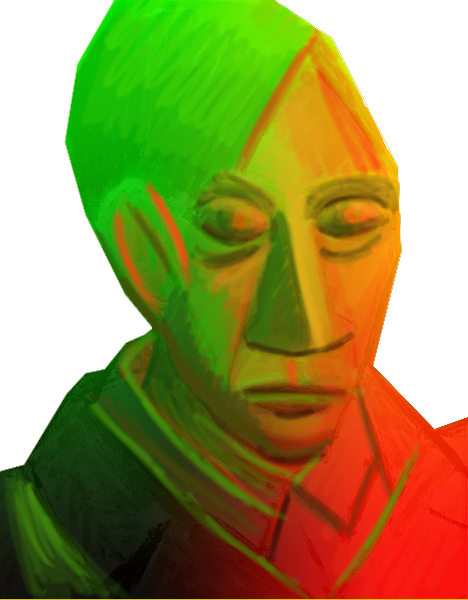}&
 \includegraphics[height=0.21\textheight]{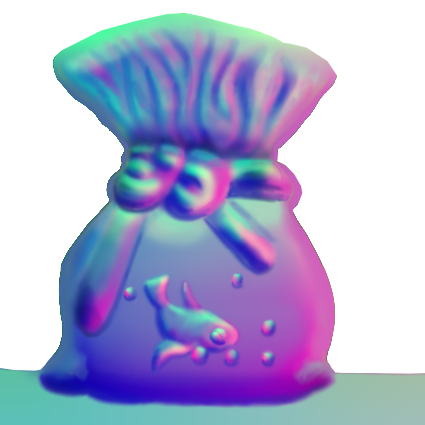}&
 \includegraphics[height=0.21\textheight]{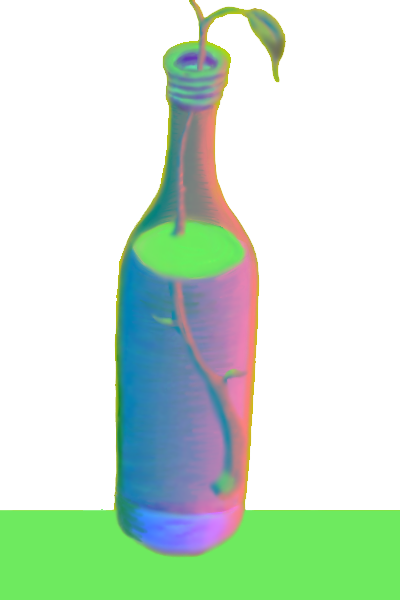}\\
\end{tabular}
\end{center}
\caption{ These shape maps are all painted manually by an artist inspired by original paintings or photographs. In all four cases, the entire process of creating a 2D vector field does not take more than one hour using a digital painting program.}
\label{fig_Shapemap1}
\end{figure}

\begin{figure}[ht]
\begin{center}
\begin{tabular}{cccccc}
 \includegraphics[height=2.10in]{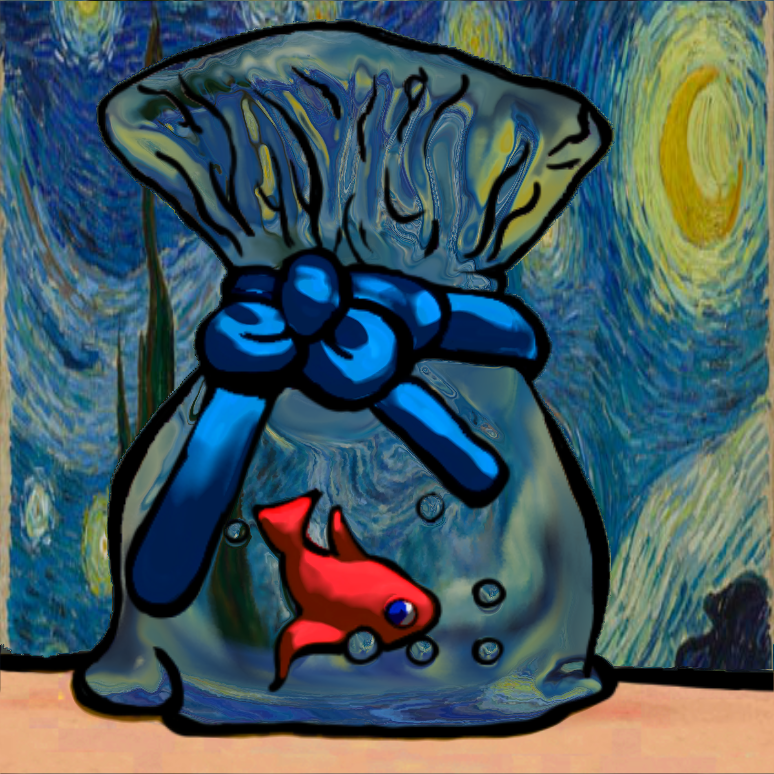}&
 \includegraphics[height=2.10in]{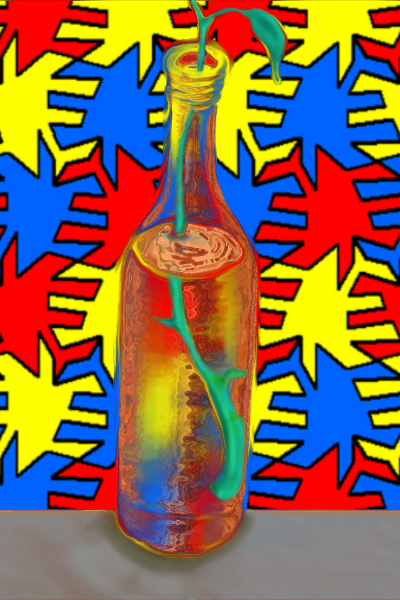}&
 \includegraphics[height=2.10in]{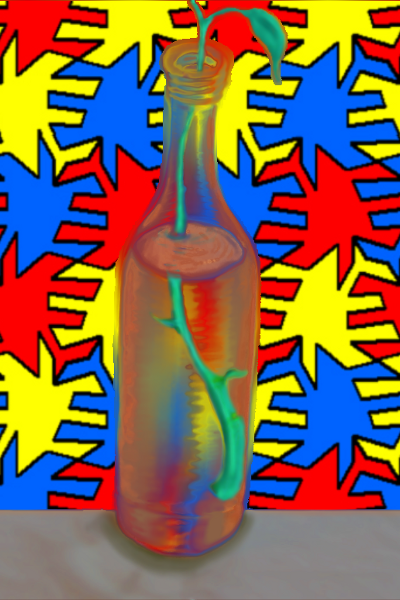}&
 \includegraphics[height=2.10in]{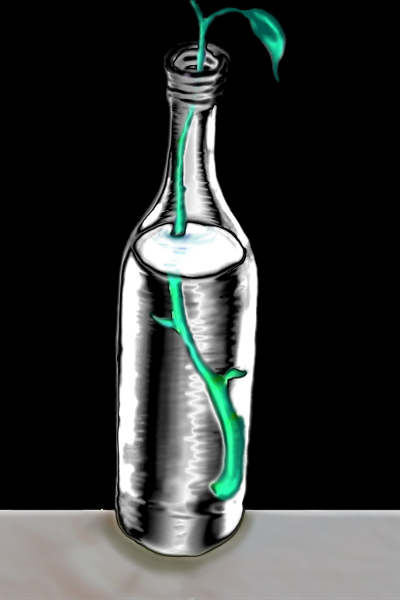}\\
  (a) Refraction \& Reflection  &
  (b) Refraction \& Reflection  &
   (c) Glossy \& Translucent   &
   (d) Fresnel term only\\
\end{tabular}
\end{center}
\caption{ Examples of non-photo-realistic compositing with reflection, glossy reflection, refraction, and translucence combined with Fresnel using hand-painted shape maps. In these examples, materials including transparency are described by a separate set of images.}
\label{fig_reflection}
\end{figure}

The thickness values $T$ are also easy to paint. $T$ values have to be nonzero for the object and zero for everywhere else. Moreover, the $T$ values have to be small close to the boundaries of the object (if considering a perspective transformation) and thin regions. This is sufficient to obtain visually correct-looking refractions. For the rest of the object, the values of $T$ can simply be any reasonable positive real number smaller than $1$. Figure~\ref{fig_Shapemap1} shows some examples of shape maps painted by artists. The thickness information is useful for artistic control of the refractions, as shown in Figure~\ref{fig_reflection}. The bottle image, in particular, shows how thickness values control the refraction. Since this bottle is half filled with a liquid, in the places where the bottle is not filled with the liquid, the $T$ value must be very small to indicate a thin glass, although the $z$ component of the unit vector is not small. 

Although we prefer artists to create shape maps, they can also convert virtual objects into shape maps. The procedure to obtain a 2D vector field is a straightforward rendering process. The $x$ and $y$ components of the 3D-normal vector of the visible point are simply converted to the red and green colors of the image. It is even possible to directly use the component $z$ of the unit normal vector as a value of $T$. However, an approximate thickness value can also be directly computed. 

\section{Photographing Objects to Create Shape Maps \label{section:PSM}}

\begin{figure}[htbp]
\begin{center}
\begin{tabular}{cccccc}
\includegraphics[height=0.2\textheight]{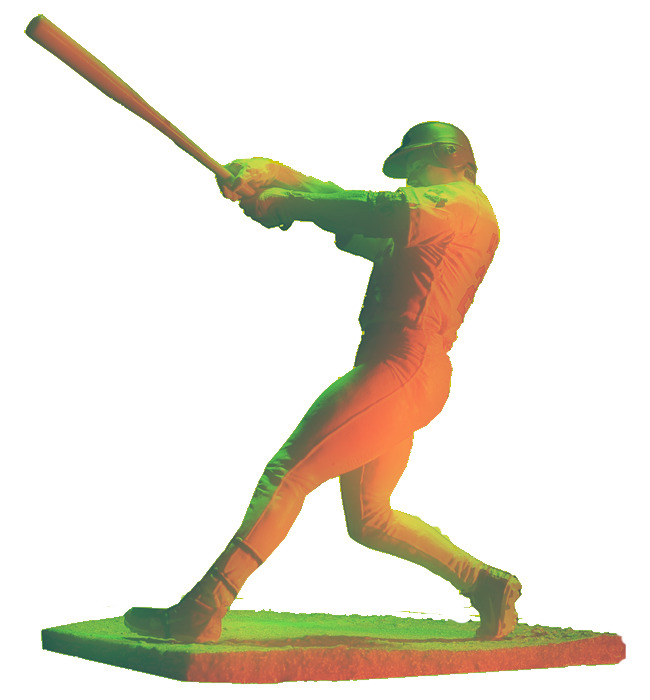}&
\includegraphics[height=0.2\textheight]{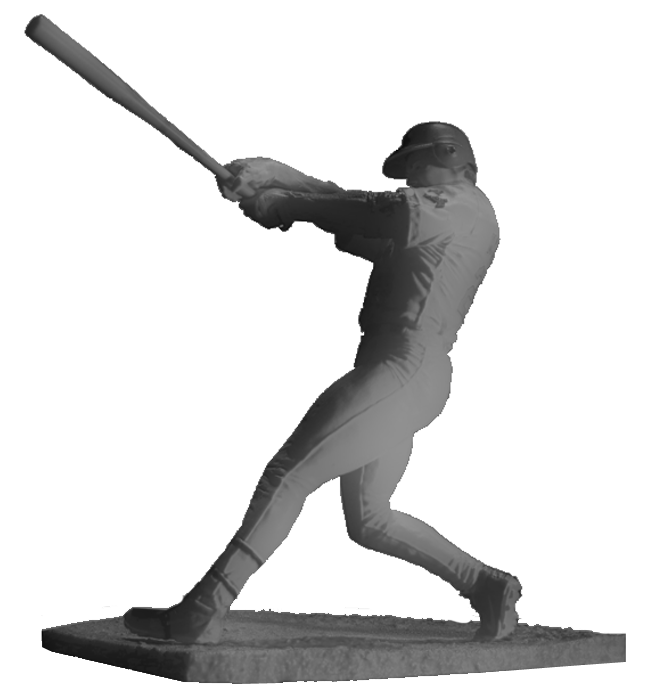}&
\includegraphics[height=0.2\textheight]{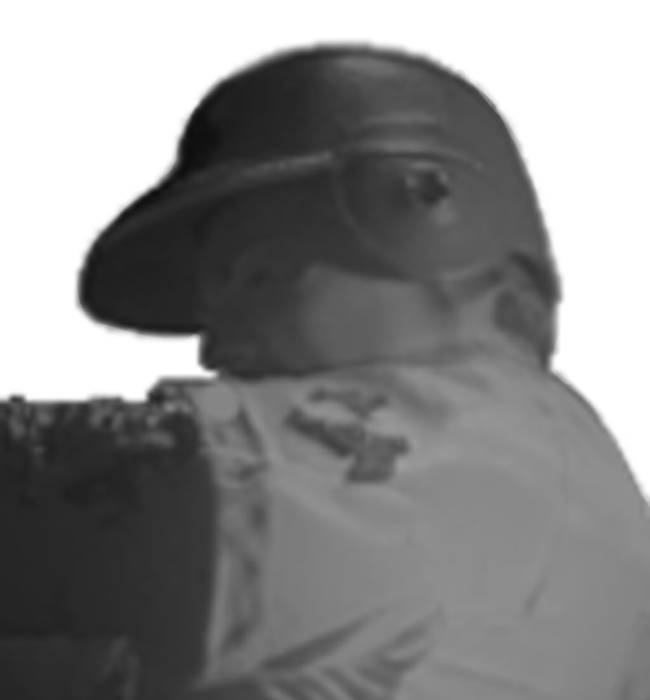}\\
\includegraphics[height=0.2\textheight]{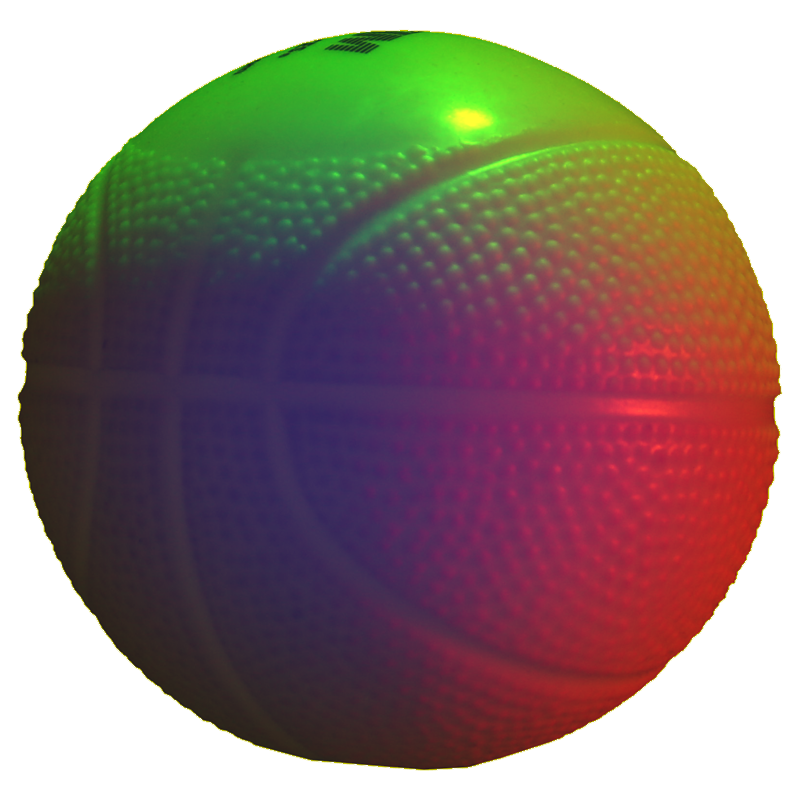}&
\includegraphics[height=0.2\textheight]{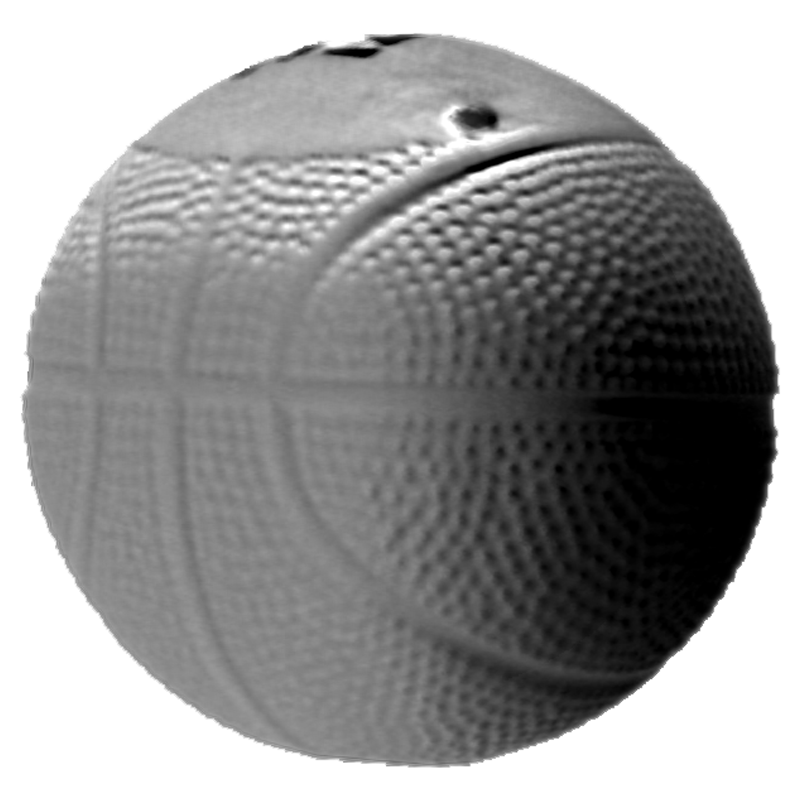}&
\includegraphics[height=0.21\textheight]{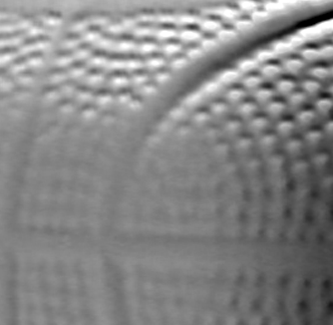}\\
\includegraphics[height=0.2\textheight]{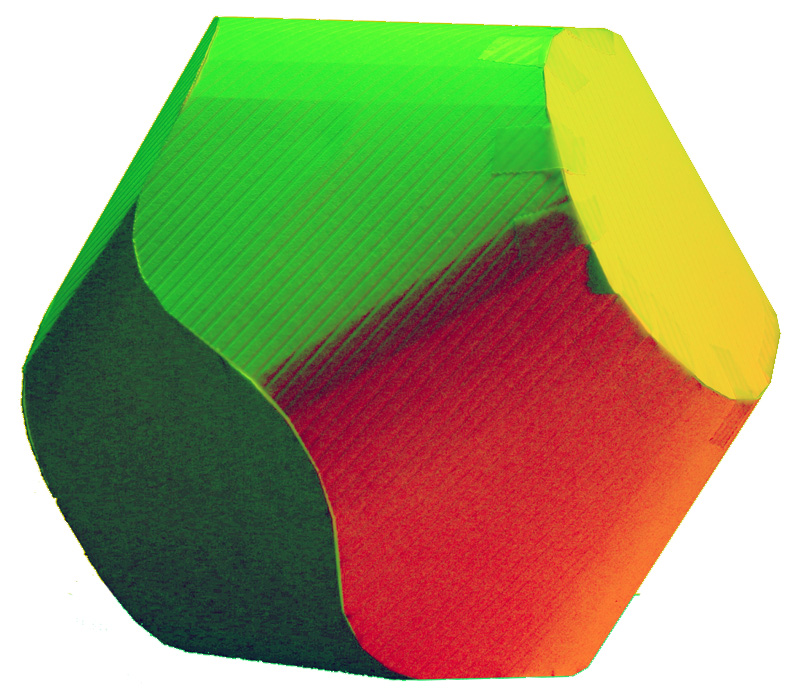}&
\includegraphics[height=0.2\textheight]{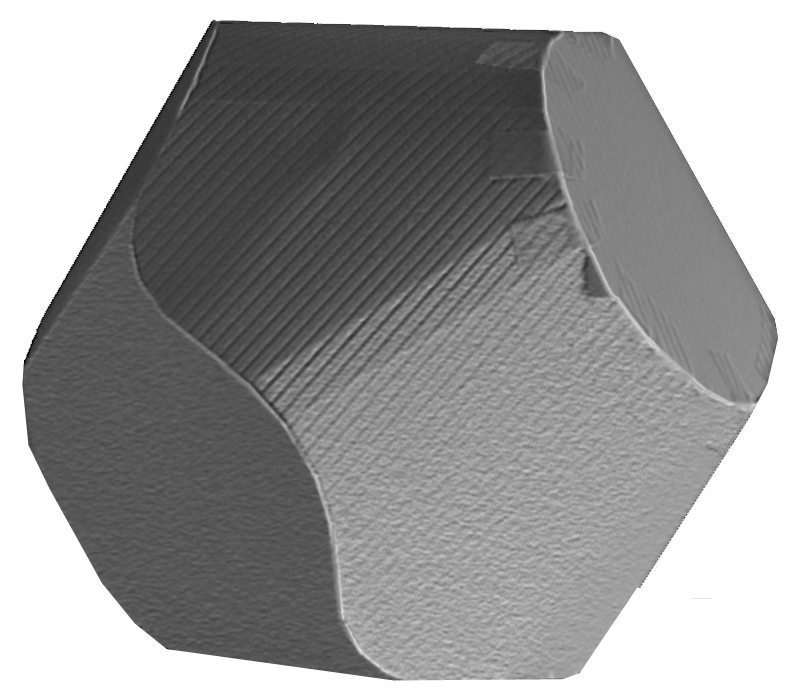}&
\includegraphics[height=0.2\textheight]{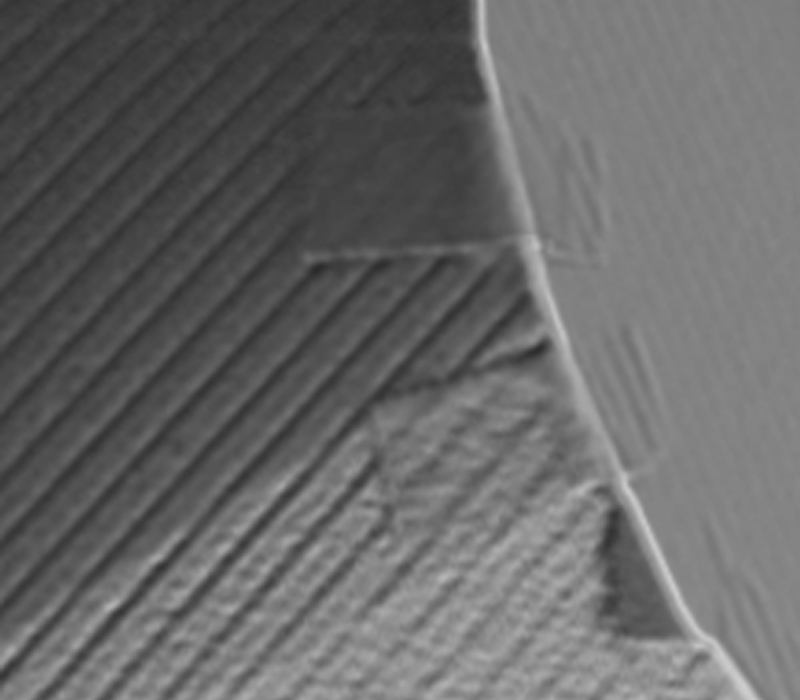}\\
 (a)  Photo shape map  &   (b) Rendering &   (c) Detail  \\
\end{tabular}
\end{center}
\caption{ Examples of shape maps generated from photographs and diffuse rendering results. As demonstrated in detailed images, it is possible to obtain unexpected visuals such as tapes that are visible in the last-detail image.}
\label{fig_shapemapphotos1}
\end{figure}

\begin{figure}[htbp]
\begin{center}
\begin{tabular}{cccccc}
\includegraphics[height=0.35\textheight]{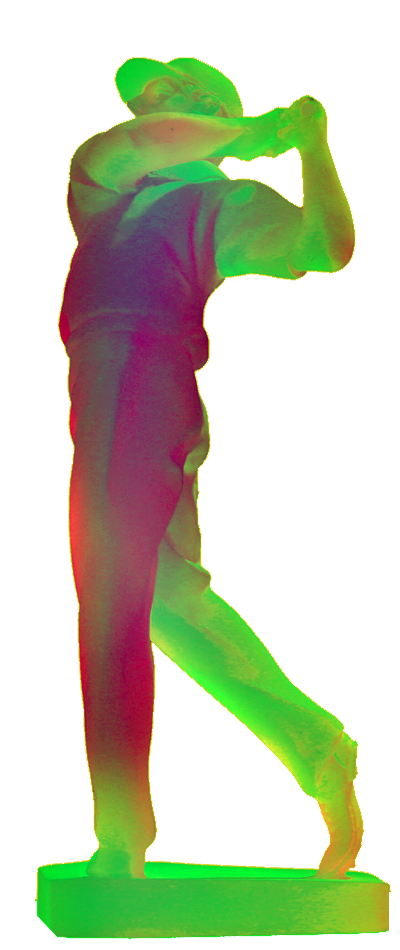}&
\includegraphics[height=0.35\textheight]{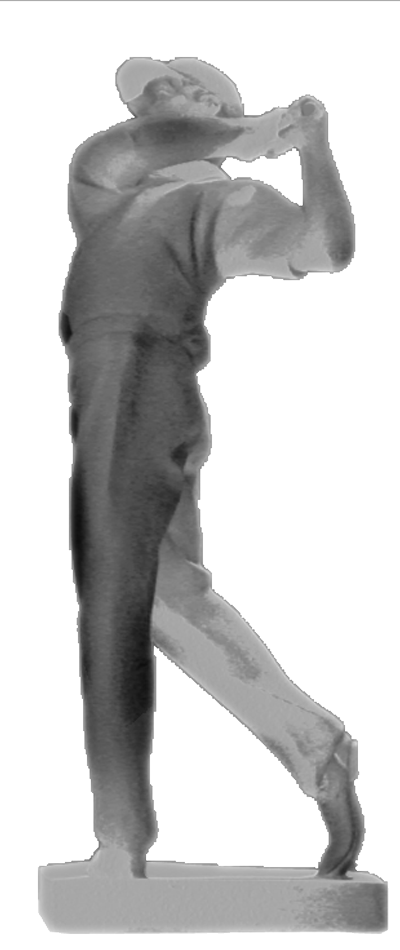}&
\includegraphics[height=0.35\textheight]{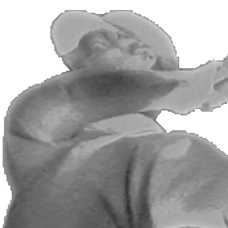}\\
\end{tabular}
\caption{ An example of a shape map generated from a photograph of translucent sculptures. The translucency of the original sculpture is visible in B\&W rendering.}
\label{fig_shapemapphotos2}
\begin{tabular}{cccccc}
\includegraphics[height=0.2\textheight]{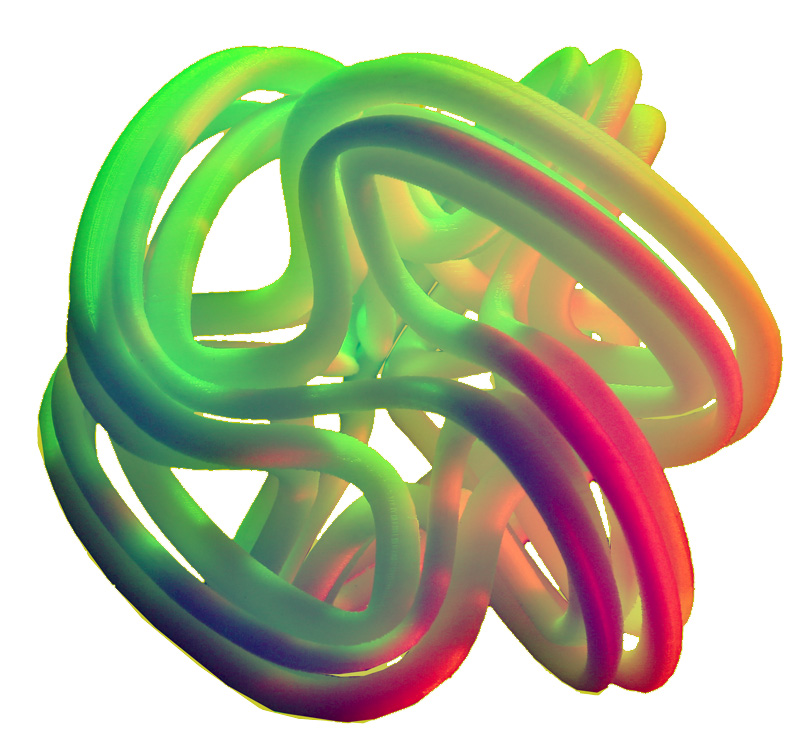}&
\includegraphics[height=0.2\textheight]{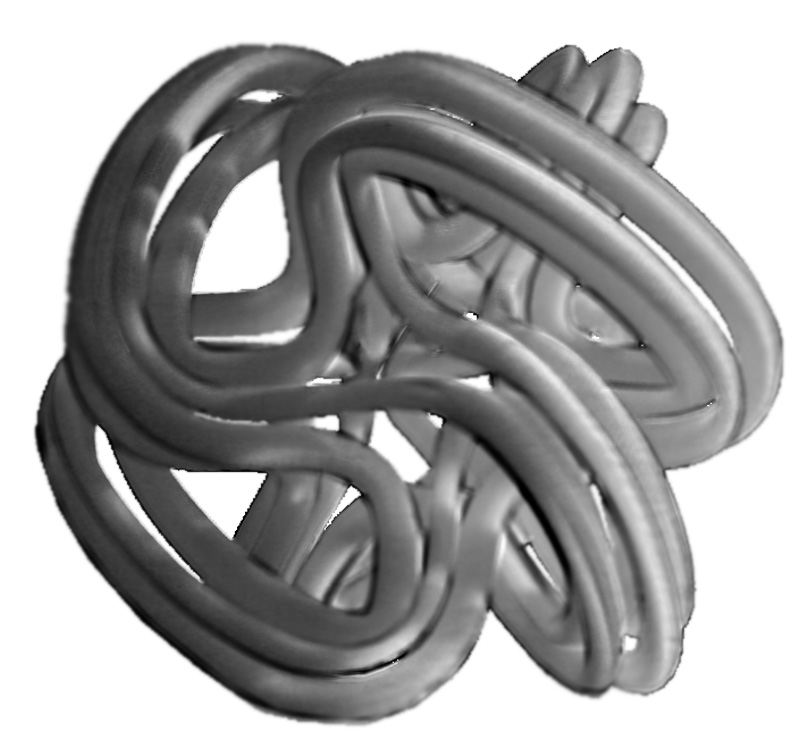}&
\includegraphics[height=0.2\textheight]{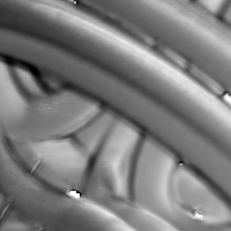}\\
\includegraphics[height=0.2\textheight]{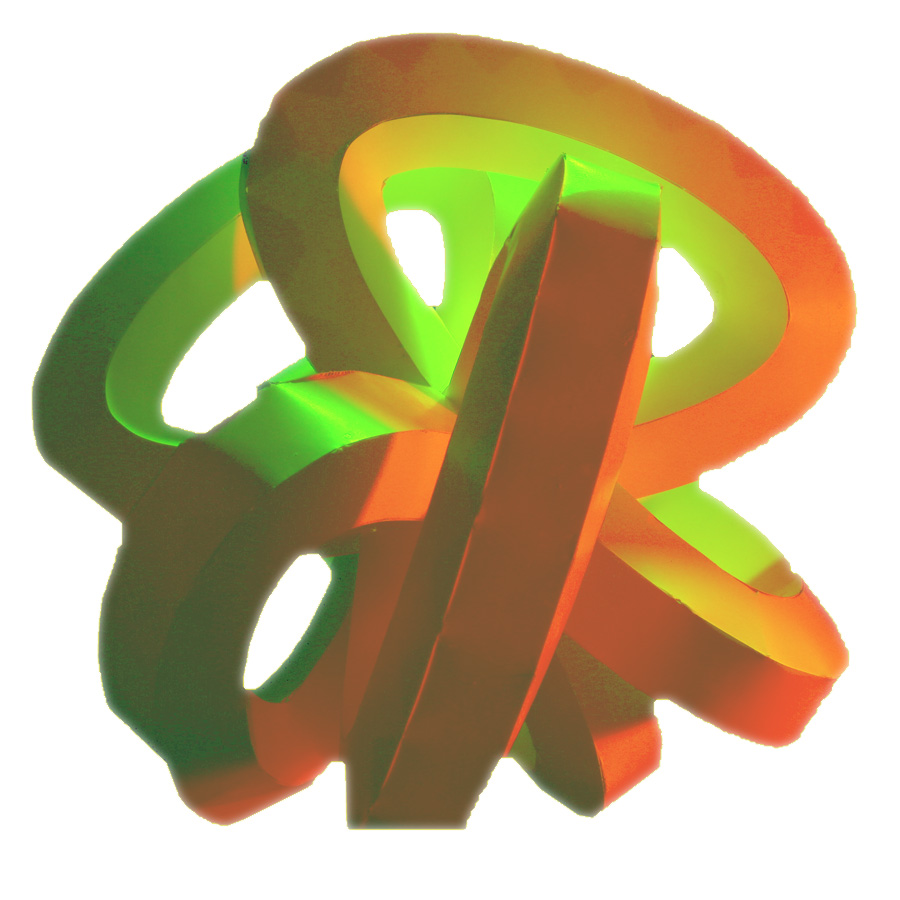}&
\includegraphics[height=0.2\textheight]{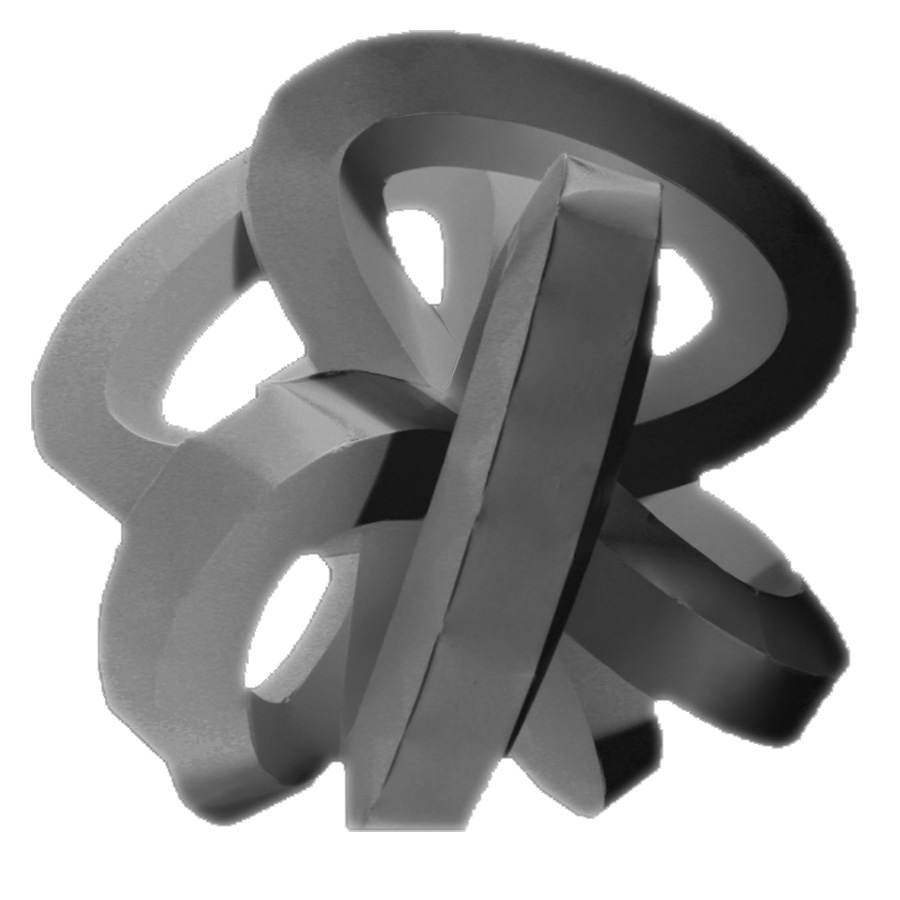}&
\includegraphics[height=0.2\textheight]{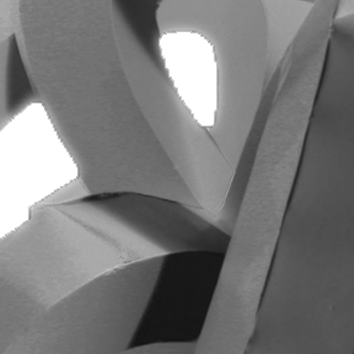}\\
 (a)  Photo shape map  &   (b) Rendering &   (c) Detail  \\
\end{tabular}
\end{center}
\caption{ Examples of shape maps generated from photographs of high genus sculptures and diffuse rendering results. As can be seen from the rendering results, the shadows created by the real lights cause some minor problems.}
\label{fig_shapemapphotos3}
\end{figure}

Note that the red and green light vectors $(1,0)$ and $(0,1)$ are linearly independent of each other. Therefore, any 2D light can be given a linear combination of the two as $(x_L,y_L)=x_L(1,0)+ y_L(0,1)$. Therefore, to compute illumination coming from an arbitrary parallel light, all we need to do is compute the contribution from two linearly independent components. This property provides another method to obtain shape maps by photographing real objects using red and green lights. This can be used as a simple alternative to environment matting \cite{Zongker1999}.  Figure~\ref{fig_shapemapphotos1}, shows such examples. As shown in the detailed images, we can even obtain minor details. One may think that this approach will not work for high-genus or transparent objects. Even in those cases, we observe that the results are unexpectedly satisfactory, as shown in Figures~\ref{fig_shapemapphotos2} and~\ref{fig_shapemapphotos3}. In these examples, we have made only minimal changes in the original image: (1) we removed and replaced backgrounds with yellow color, and (2) we added a nonzero blue value for object regions. Although a constant thickness is not correct, the resulting refractions, which are not shown in these examples, appear reasonably convincing. Artists, of course, can further manipulate these photographs to obtain the desired effects.

\section{Illustrating/Sketching Shape Maps \label{section:ISM}}

\begin{figure}[ht]
\begin{center}
\begin{tabular}{cccccc}
 \includegraphics[width=0.32\textwidth]{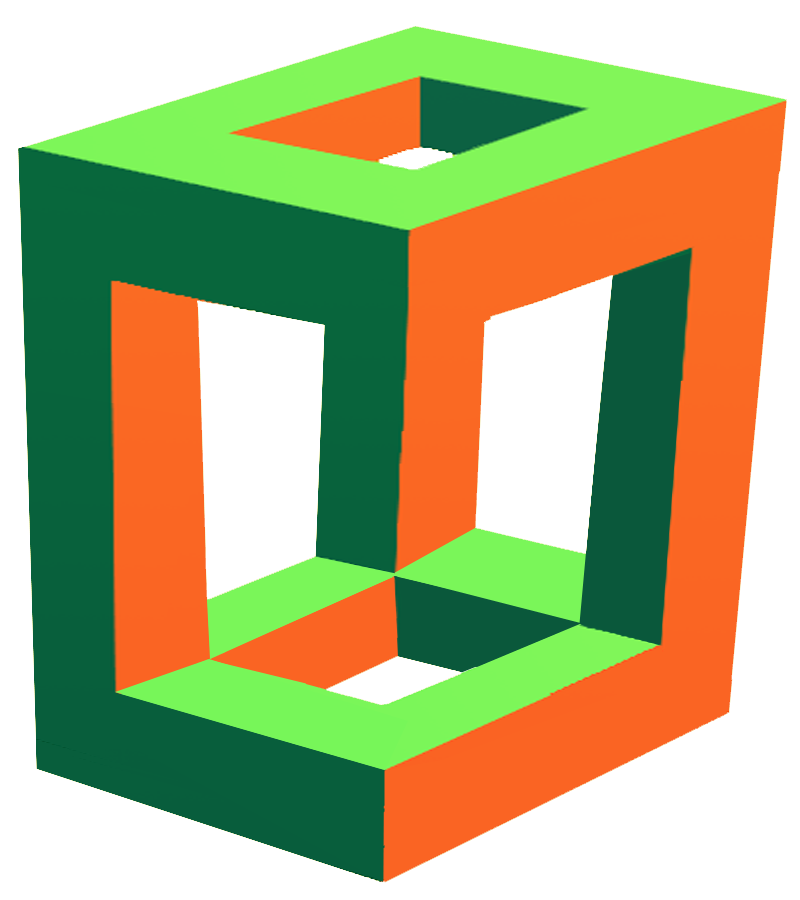}&
\includegraphics[width=0.32\textwidth]{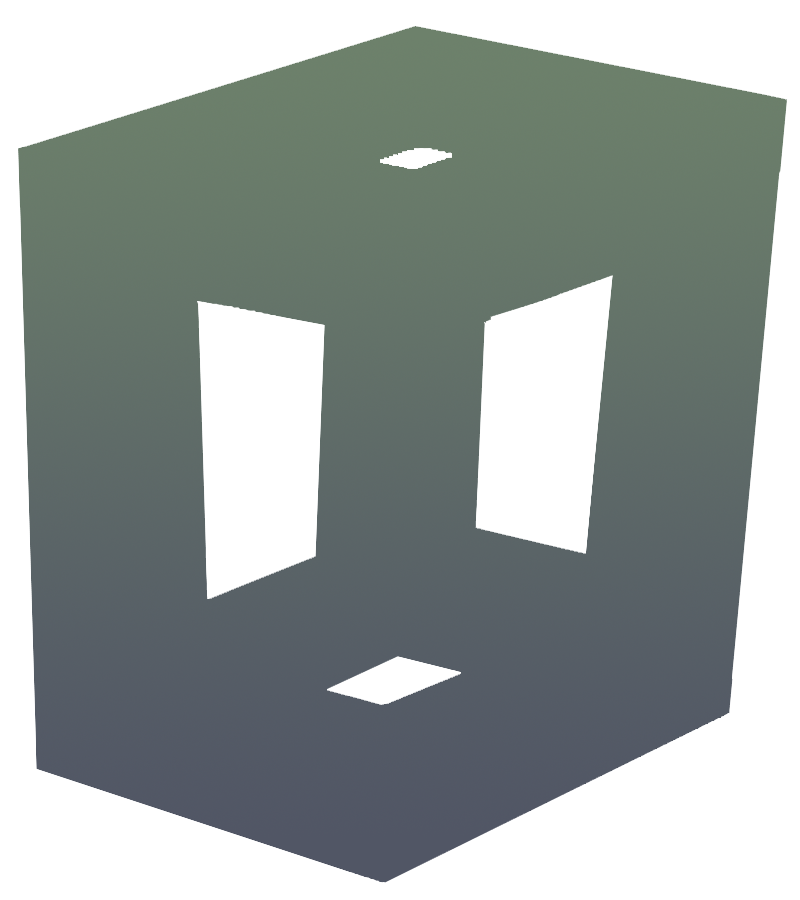}&
\includegraphics[width=0.32\textwidth]{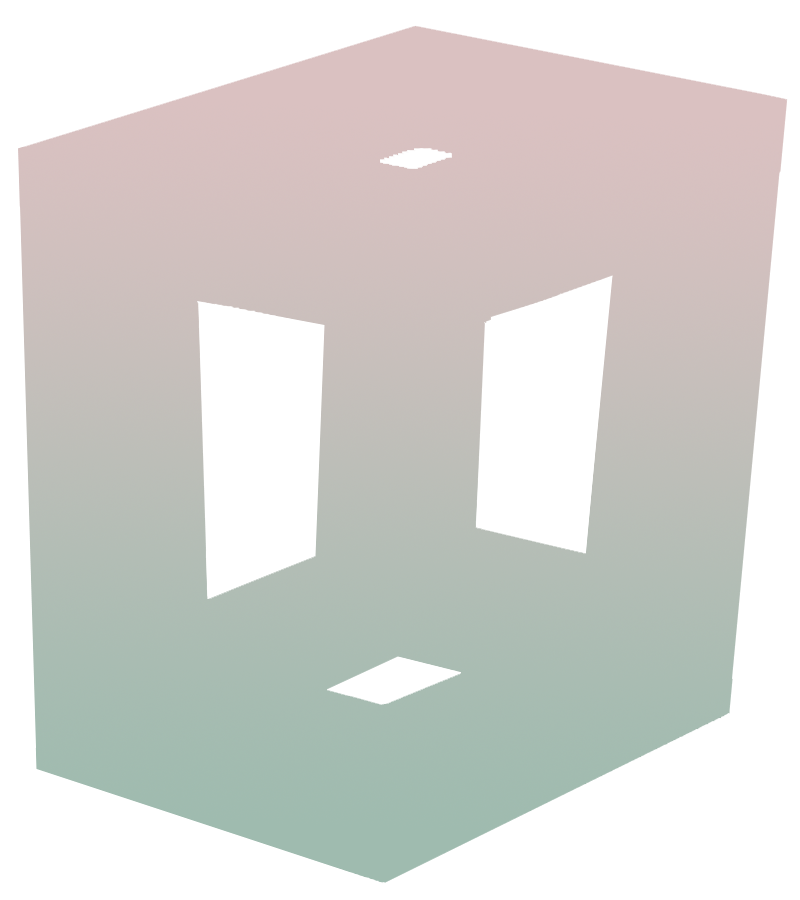}\\
\end{tabular}
\end{center}
\caption{ An example of an impossible object. The shape map is created using our program. The other two images are parameters of a barycentric shader. These three images provide shape and material representations of the impossible object.}
\label{fig_impossible3_rep}
\end{figure}

\begin{figure}[ht]
\begin{center}
\begin{tabular}{cccccc}
 \includegraphics[width=0.32\textwidth]{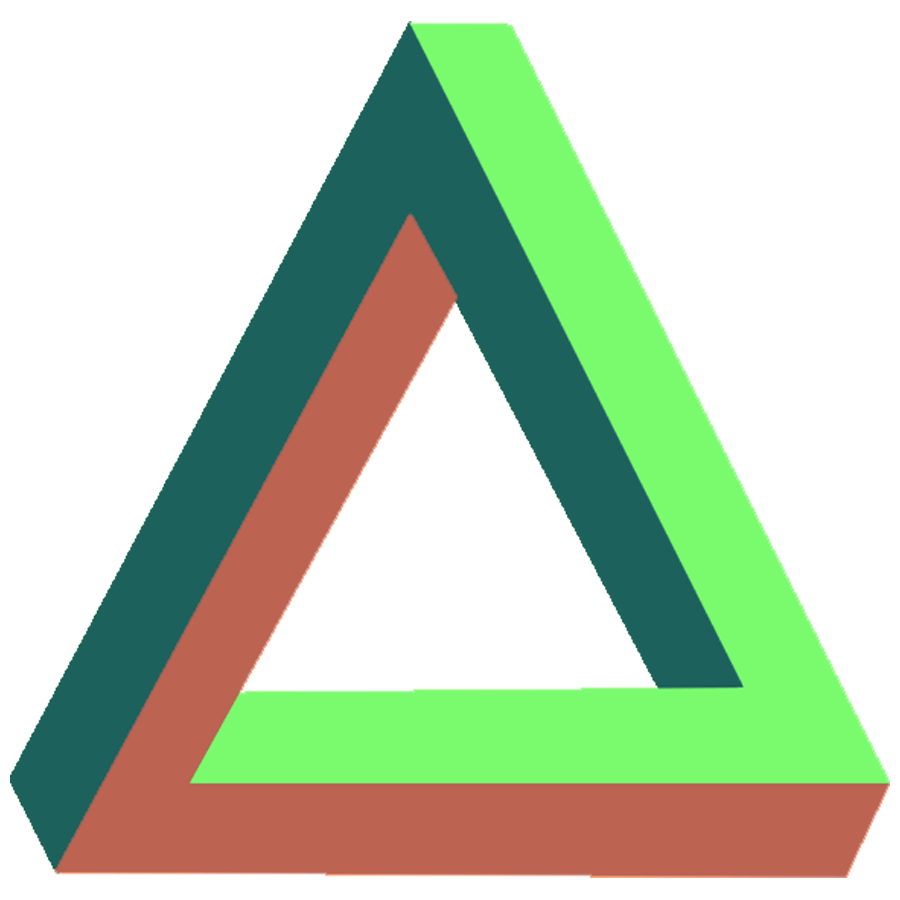}&
\includegraphics[width=0.32\textwidth]{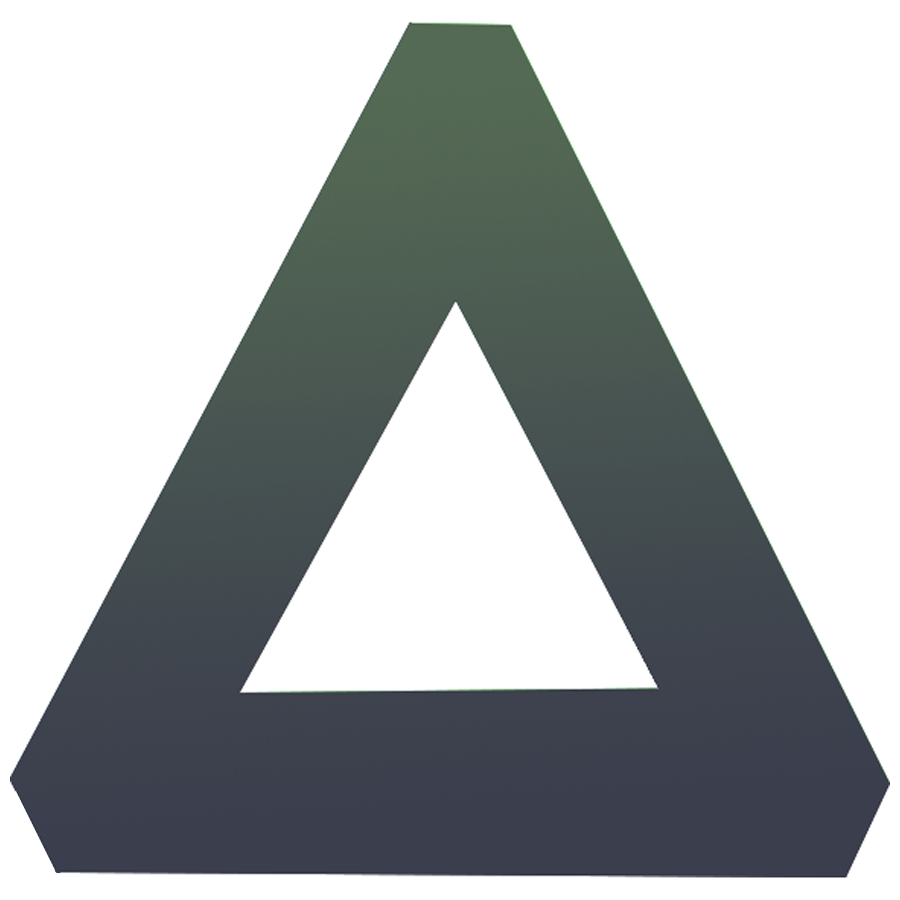}&
\includegraphics[width=0.32\textwidth]{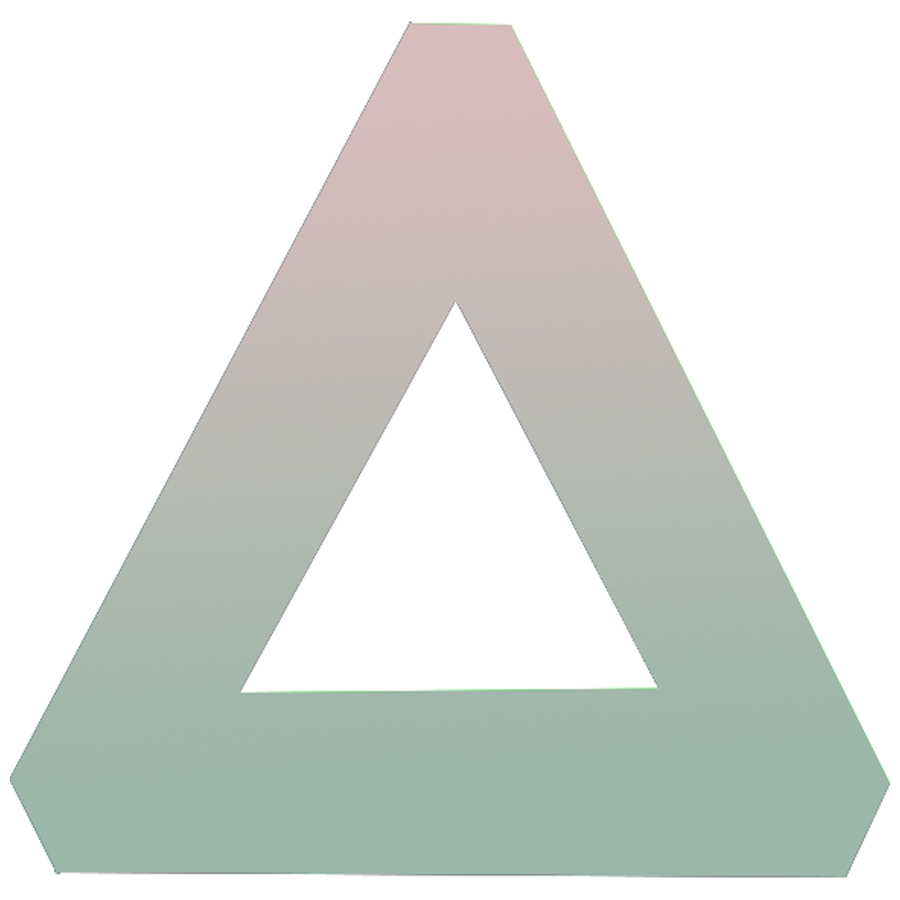}\\
\end{tabular}
\end{center}
\caption{ An example of an impossible object. The shape map is created using our program. The other two images are parameters of a barycentric shader. These three images provide shape and material representations of the impossible object.}
\label{fig_impossible1_rep}
\end{figure}

The most viable option to create shape maps is to model 2D vector fields directly with a sketch-based interface. As discussed earlier, there already exist many sketch-based interface approaches, such as those of Lumo \cite{Johnston2002} or CrossShade \cite{Shao2012}, that can be used directly to create normal maps as shape maps, as shown in \ref{fig_Normal}. To construct non-conservative vector fields, however, there is a need to provide more control to users. We have developed a sketch-based integrated mock-3D scene modeling system that can allow users to obtain any vector field. For example, we have created impossible objects shown in Figures~\ref{fig_impossible3_rep}, and ~\ref{fig_impossible1_rep} using our system. 

\begin{figure}[ht]
\begin{center}
\begin{tabular}{cccccc}
 \includegraphics[width=0.32\textwidth]{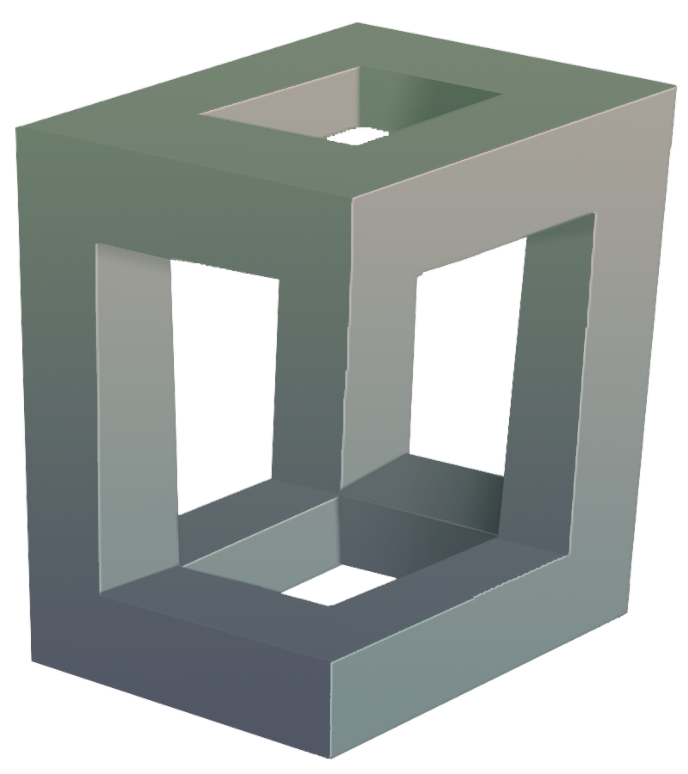}&
 \includegraphics[width=0.32\textwidth]{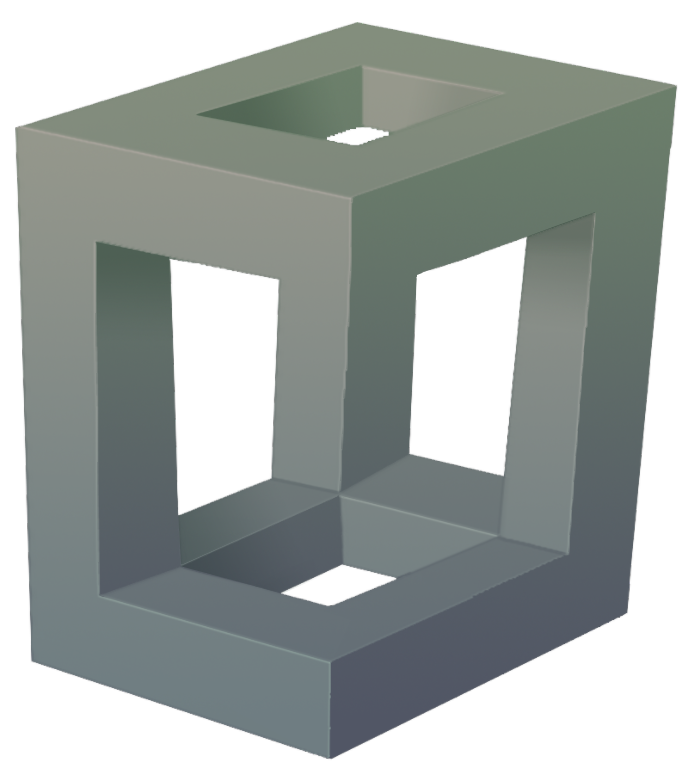}&
  \includegraphics[width=0.32\textwidth]{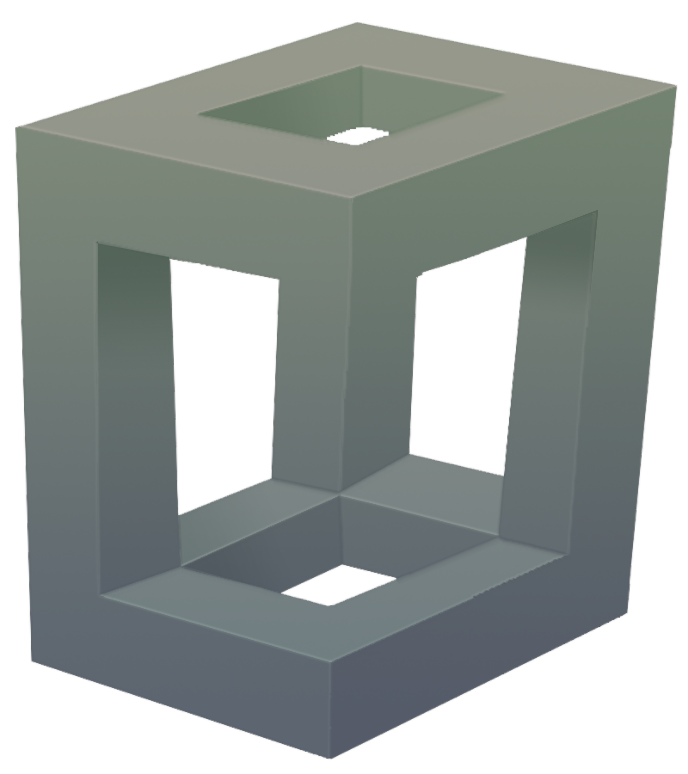}\\
\end{tabular}
\end{center}
\caption{ Diffuse renderings of the impossible object in Figure~\ref{fig_impossible3_rep} with ambient occlusion and local shadows. Note how much visual quality improves with subtle local shadows and beveled-edge look caused by ambient occlusion that is obtained from dynamically computed geometry information. }
\label{fig_impossible3_render}
\end{figure}

In our system, we represent mock-3D shapes using 2-manifold quadrilateral meshes with boundaries. Each quadrilateral face of the manifold mesh is a cubic Bezier patch based on the tensor product \cite{beatty1987}. We choose parametric formulation over subdivision to allow, in particular, valent-2 vertices. With cubic patches, users can easily obtain $G^1$ continuity or introduce derivative discontinuities. We construct pseudo-2-complexes by stitching the boundaries of these quad-meshes. Internally, these meshes are still kept as 2-manifold surfaces. To obtain boundaries, we simply label some faces as "invisible". The advantage of this flexibility is that we can design arbitrary complex manifold structures. To handle curved edges, we extended an existing manifold data structure to include cubic Bezier curves as edge shapes. 

\begin{figure}[ht]
\begin{center}
\begin{tabular}{cccccc}
 \includegraphics[width=0.32\textwidth]{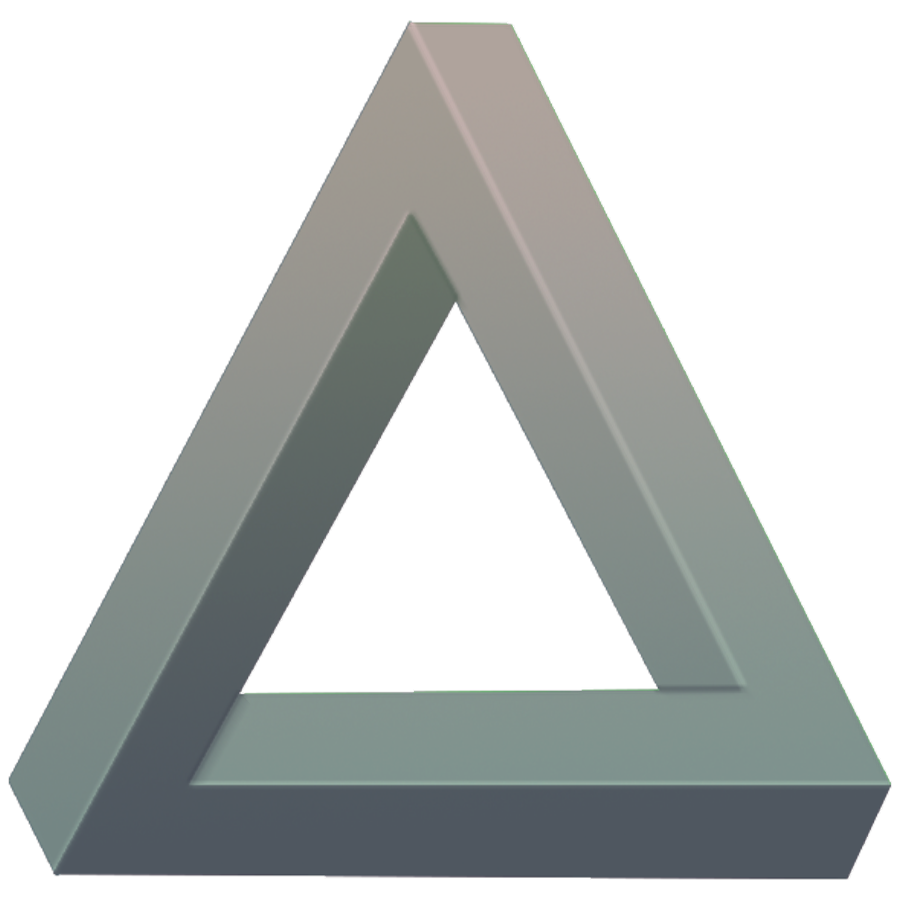}&
 \includegraphics[width=0.32\textwidth]{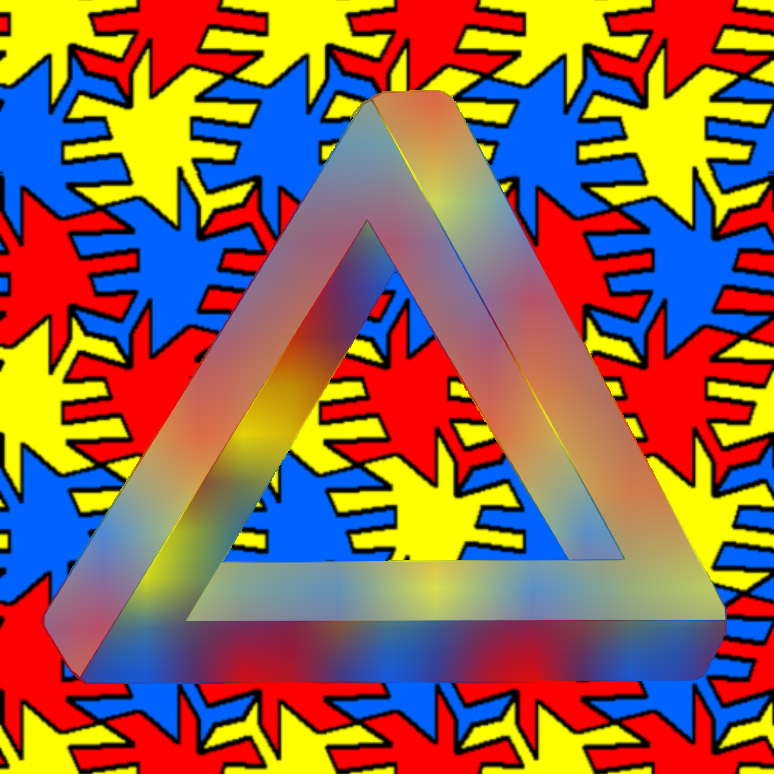}&
  \includegraphics[width=0.32\textwidth]{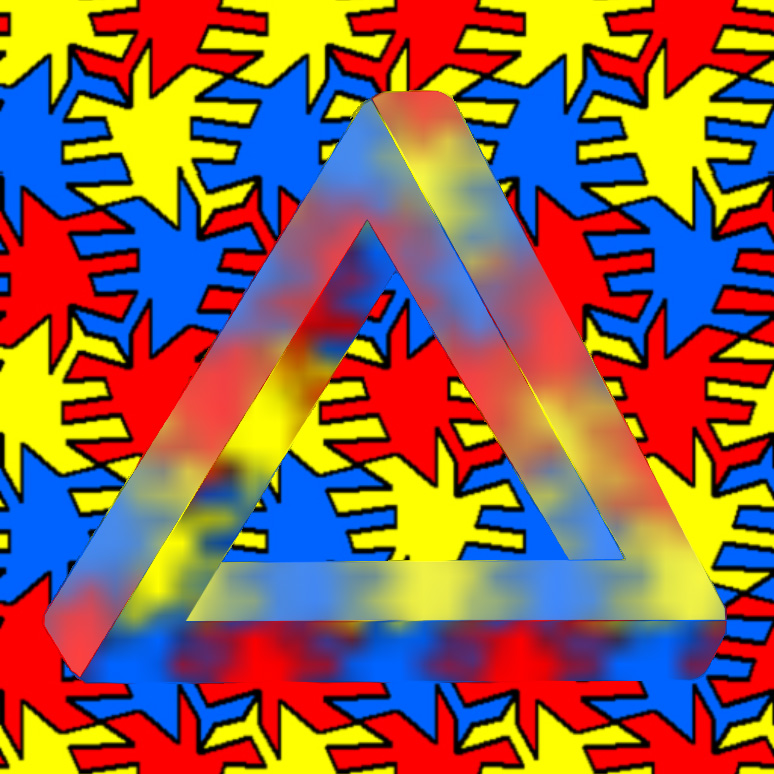}\\
\end{tabular}
\end{center}
\caption{ Diffuse and Transparent renderings of the impossible object in Figure~\ref{fig_impossible1_rep}. Note that since we have thickness information, we can also obtain realistic-looking transparency and translucency with impossible objects. }
\label{fig_impossible1_render}
\end{figure}

The system looks and feels exactly like a 2D vector graphics system, in which users can only change the 2D positions of control points. On the other hand, we allow users to change the $z$ positions of control points. Note that the endpoints of the Bezier control points correspond to the corners of the vertices of the manifold meshes. In other words, it is possible to have $G^0$ discontinuities in $z$ positions in the boundaries of two neighboring patches while visually looking stitched together.

\begin{figure}[ht]
\begin{tabular}{cccccc}
\includegraphics[width=0.45\textwidth]{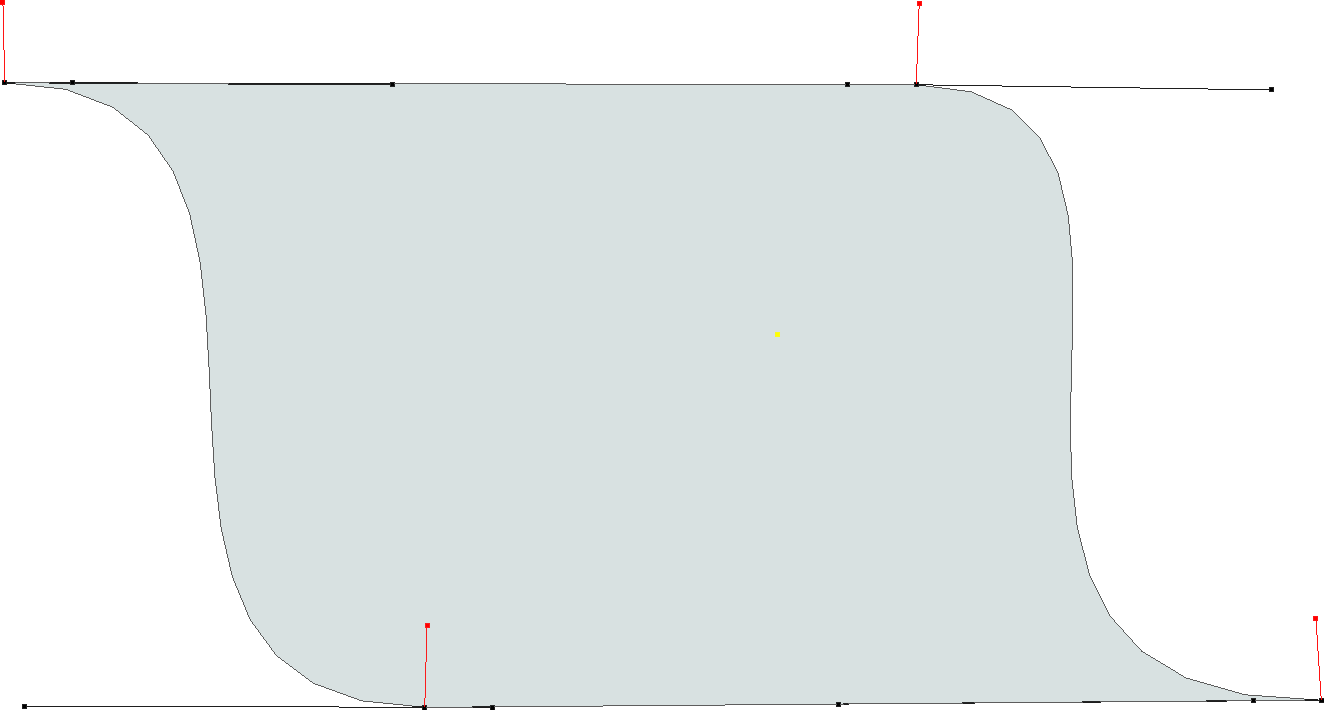}&
\includegraphics[width=0.45\textwidth]{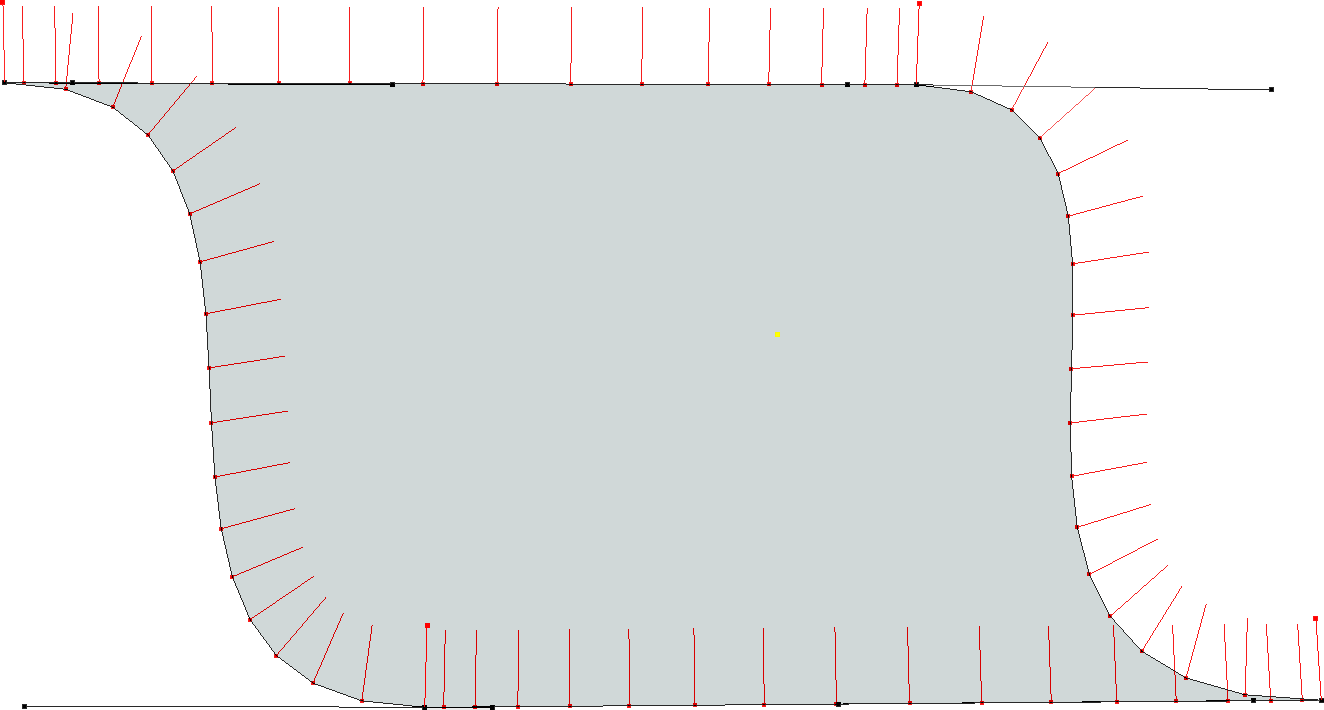}\\
 (a) Control vectors&
  (b) Boundary vector Interpolation \\
\includegraphics[width=0.45\textwidth]{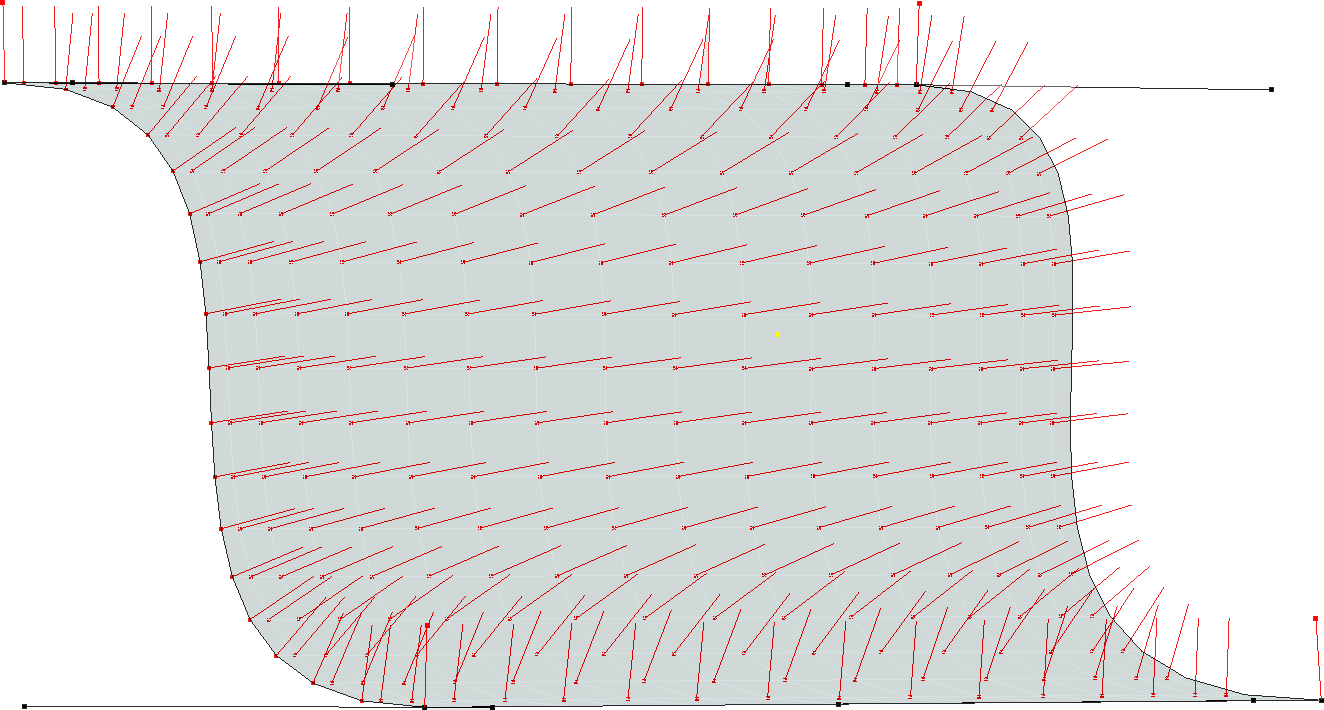}&
\includegraphics[width=0.45\textwidth]{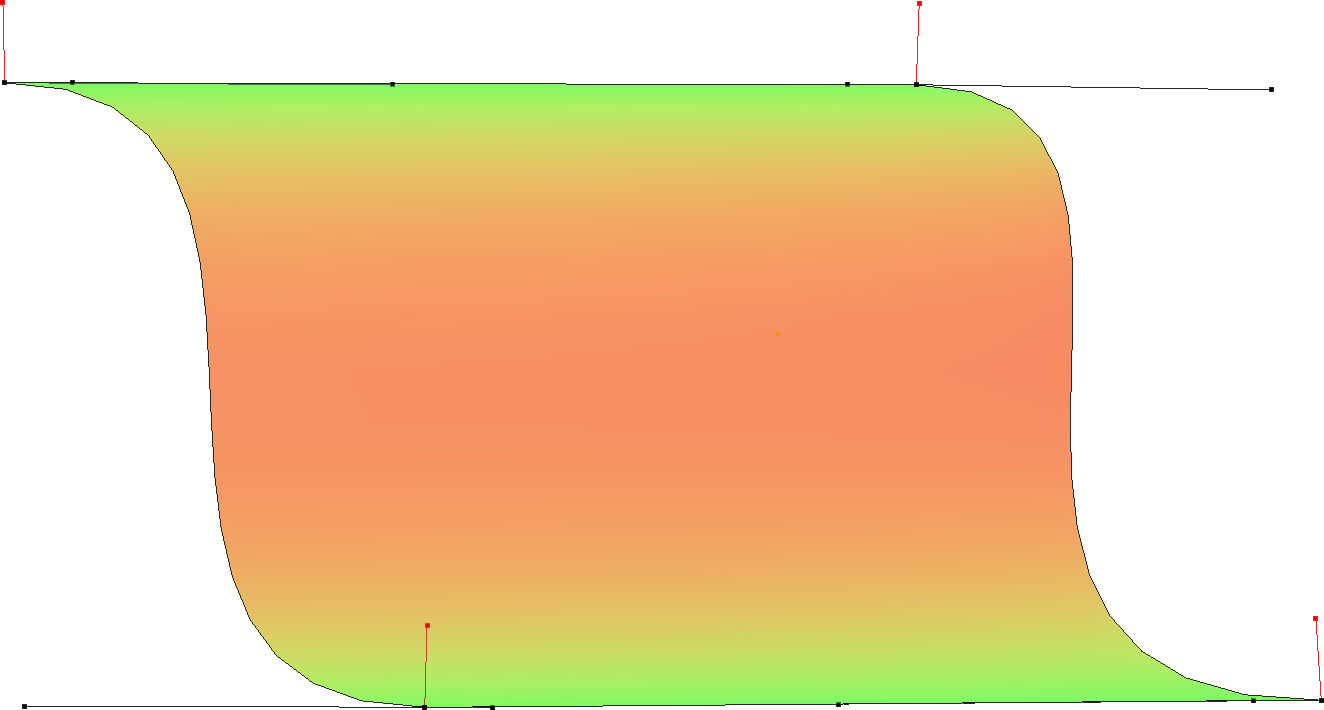}\\
   (c) Coons Interpolation &
  (d) Shape map  \\
\end{tabular}
\caption{   An example of shape map creation on a Bezier patch by interpolating control vectors first along the edges, then inside the patch. The thickness is constant}
\label{fig_vectorfield}
\end{figure} 

\begin{figure}[ht]
\begin{tabular}{cccccc}
\includegraphics[width=0.45\textwidth]{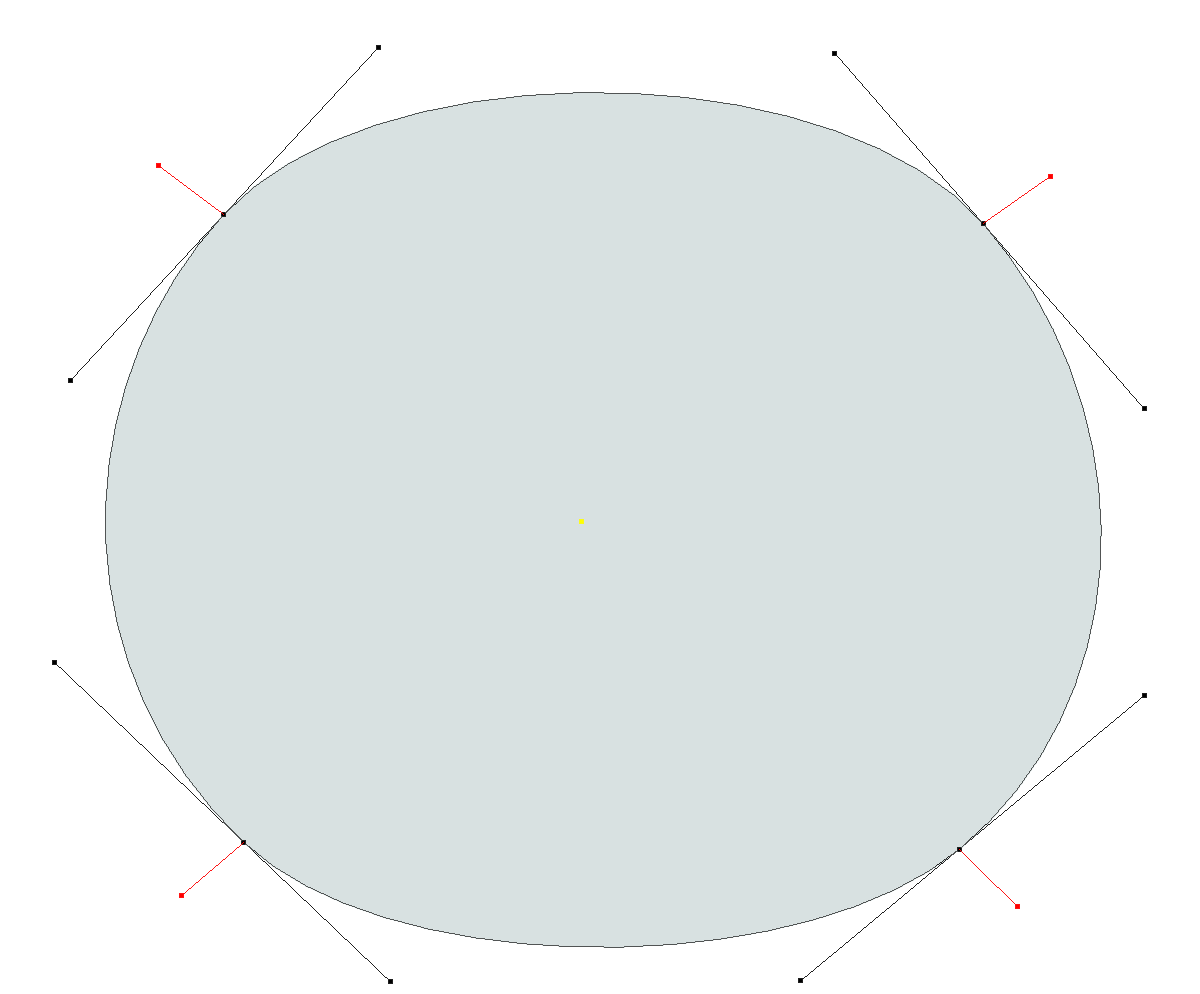}&
\includegraphics[width=0.45\textwidth]{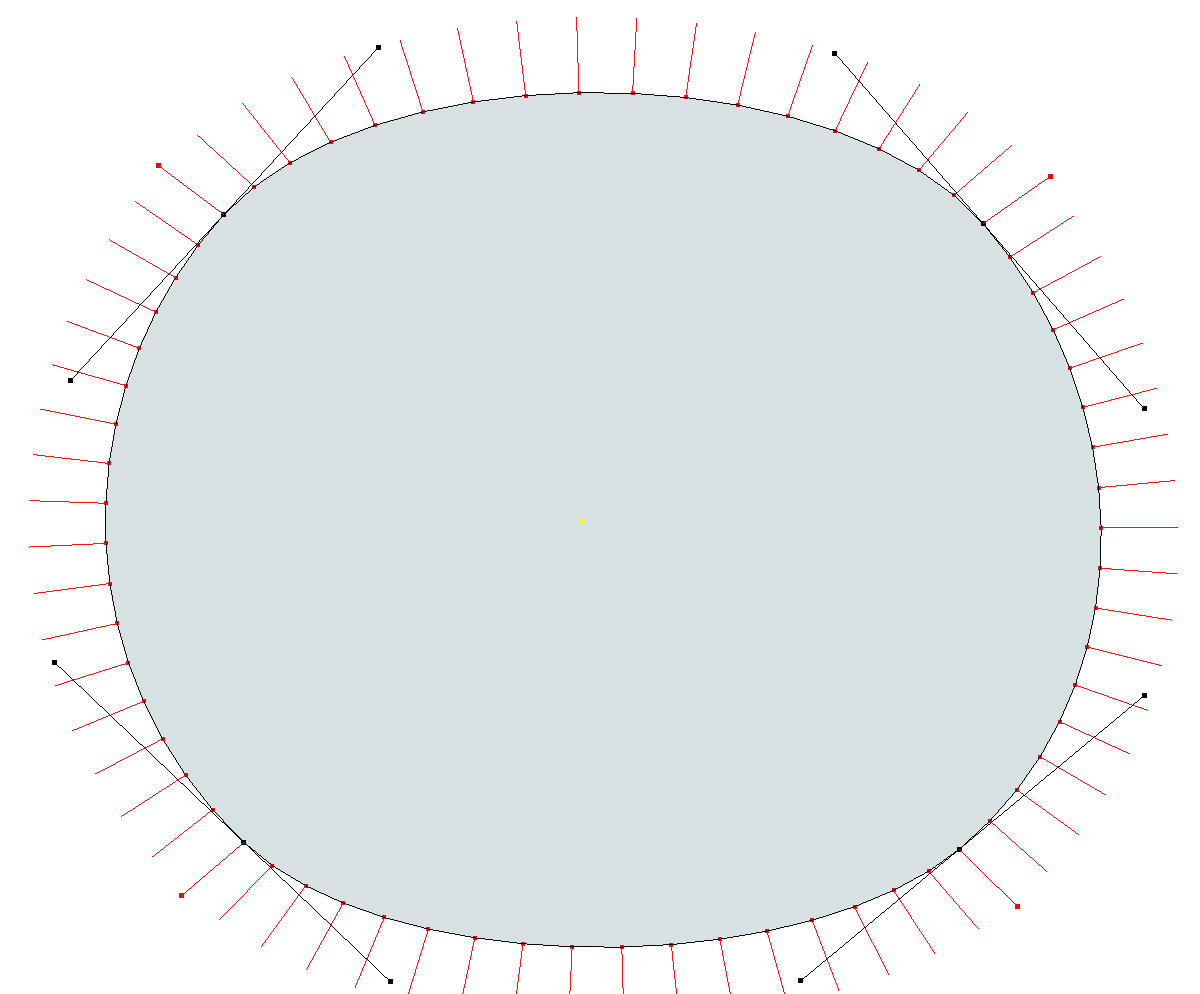}\\
 (a) Control vectors&
  (b) Boundary vector Interpolation \\
\includegraphics[width=0.45\textwidth]{vectorfield/a3}&
\includegraphics[width=0.45\textwidth]{vectorfield/a4}\\
   (c) Coons Interpolation &
  (d) Shape map  \\
\end{tabular}
\caption{   An example of gradient field (conservative vector field) creation on a Bezier patch by interpolating control vectors first along the edges, then inside the patch. The thickness is constant}
\label{fig_vectorfield_a}
\end{figure} 

\begin{figure}[ht]
\begin{tabular}{cccccc}
\includegraphics[width=0.45\textwidth]{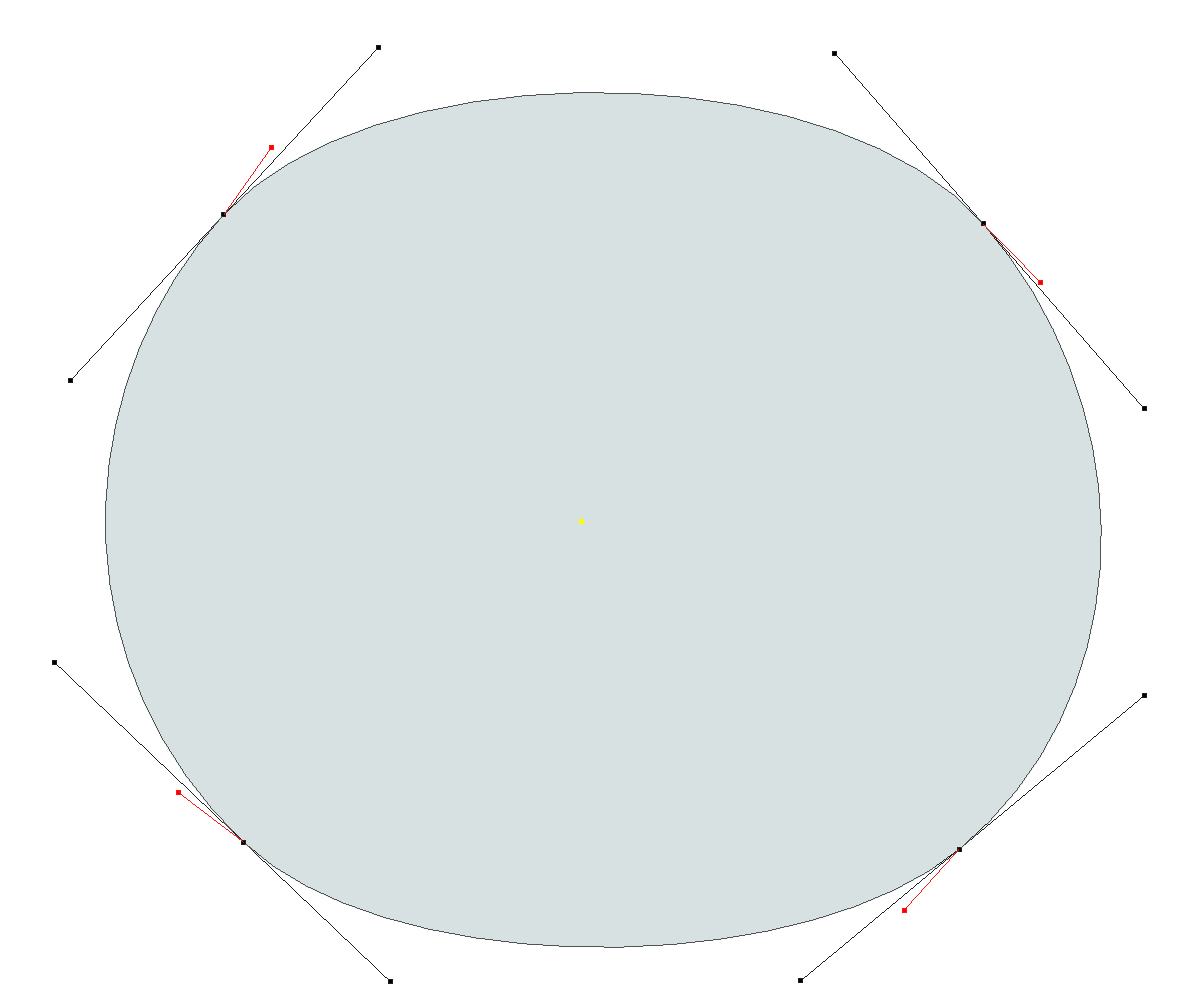}&
\includegraphics[width=0.45\textwidth]{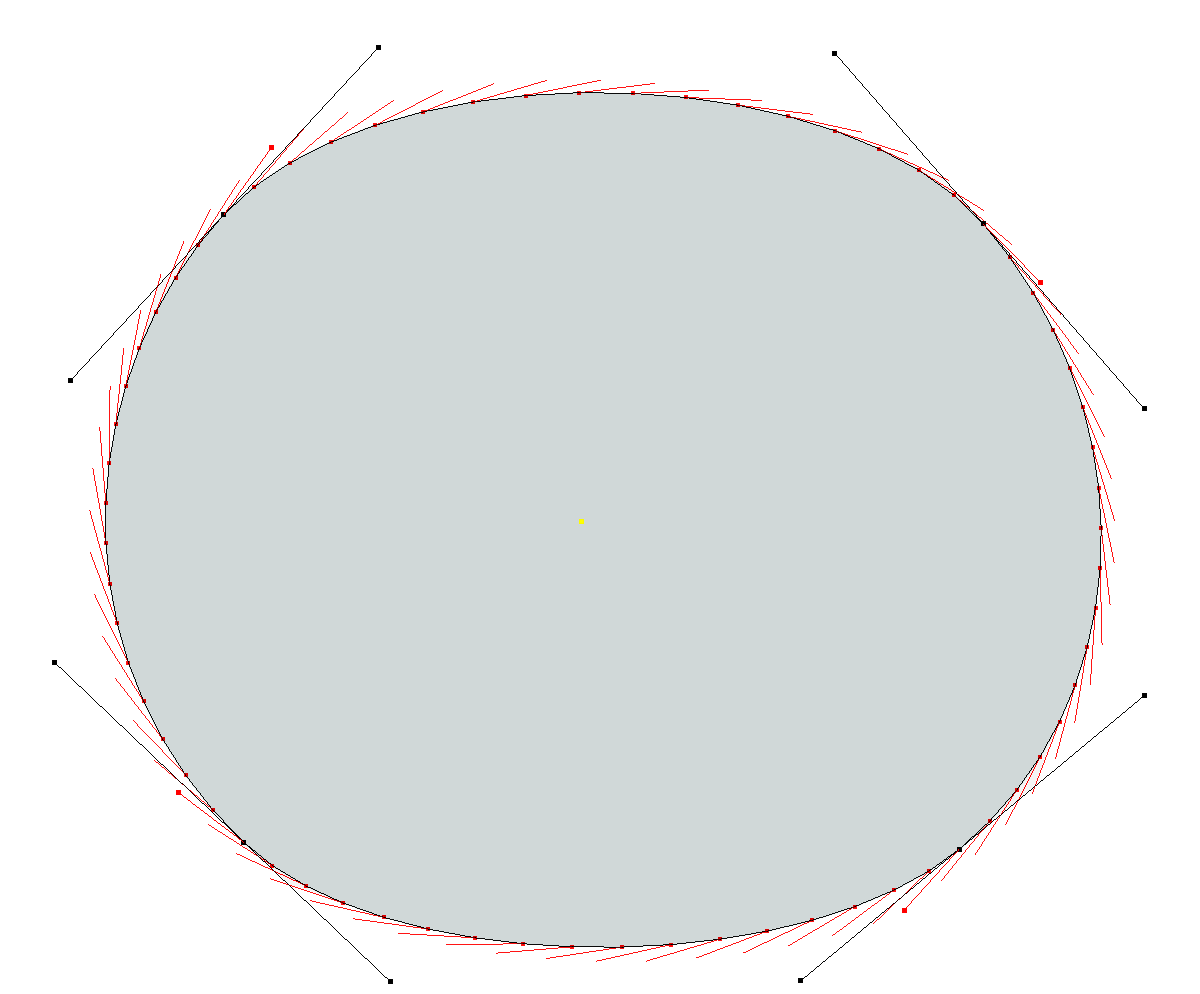}\\
 (a) Control vectors&
  (b) Boundary vector Interpolation \\
\includegraphics[width=0.45\textwidth]{vectorfield/c3}&
\includegraphics[width=0.45\textwidth]{vectorfield/c4}\\
   (c) Coons Interpolation &
  (d) Shape map  \\
\end{tabular}
\caption{   An example of non-conservative vector field creation on a Bezier patch by interpolating control vectors first along the edges, then inside the patch. The thickness is constant}
\label{fig_vectorfield_c}
\end{figure} 
 
The quad-patch-based representation provides simple control of 2D vector fields. In our system, at each corner, a 2D vector is assigned. Initial --default-- assignments are created using boundary gradient information \cite{Johnston2002} and local normals \cite{Shao2012}. Users can change these default assignments simply by changing the 2D vector, as shown in Figures~\ref{fig_vectorfield}(a),~\ref{fig_vectorfield_a}(a), and~\ref{fig_vectorfield_c}(a). We first compute the 2D vector field along the curved edges by rotating the vectors along the curves, as shown in Figures~\ref{fig_vectorfield}(b),~\ref{fig_vectorfield_a}(b), and~\ref{fig_vectorfield_c}(b).  These 1D vector fields defined at the edges serve as boundary functions that can later be interpolated using Coons patches to fill the inside of quad patches \cite{beatty1987}, as shown in Figure~\ref{fig_vectorfield}(c),~\ref{fig_vectorfield_a}(c), and~\ref{fig_vectorfield_c}(c). Then, converting it into a shape map is a simple process, as in Figures~\ref{fig_vectorfield}(d),~\ref{fig_vectorfield_a}(d), and~\ref{fig_vectorfield_c}(d). Figures~\ref{fig_vectorfield_a} and~\ref{fig_vectorfield_c} show how conservative and non-conservative fields can be obtained inside the same Bezier patch just by changing the direction of the control vector. We chose quadrilaterals as the main structure, since a straightforward Coons interpolation formula exists only for quadrilaterals. Therefore, our system only provides operations to create quadrilaterals. The user can also create triangles since they can be included by making one edge zero length. 

\begin{figure}[htbp]
\begin{center}
\begin{tabular}{cccccc}
\includegraphics[width=0.32\textwidth]{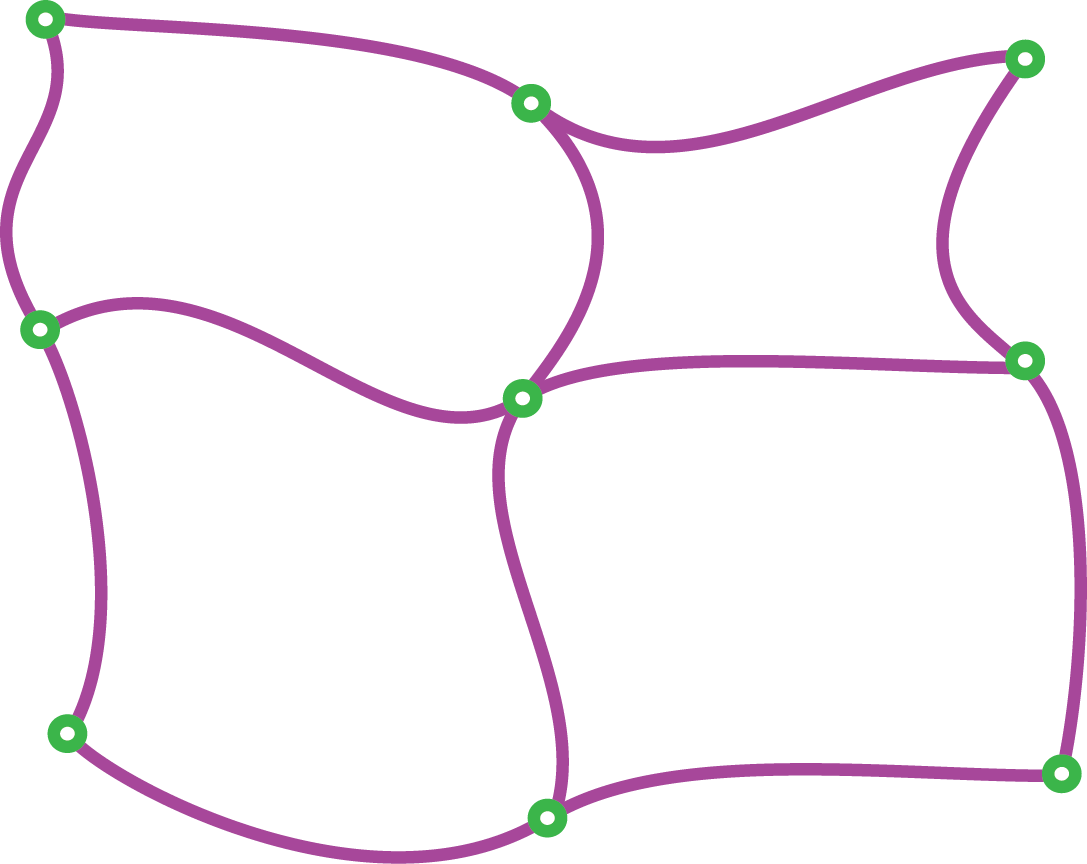}&
\includegraphics[width=0.32\textwidth]{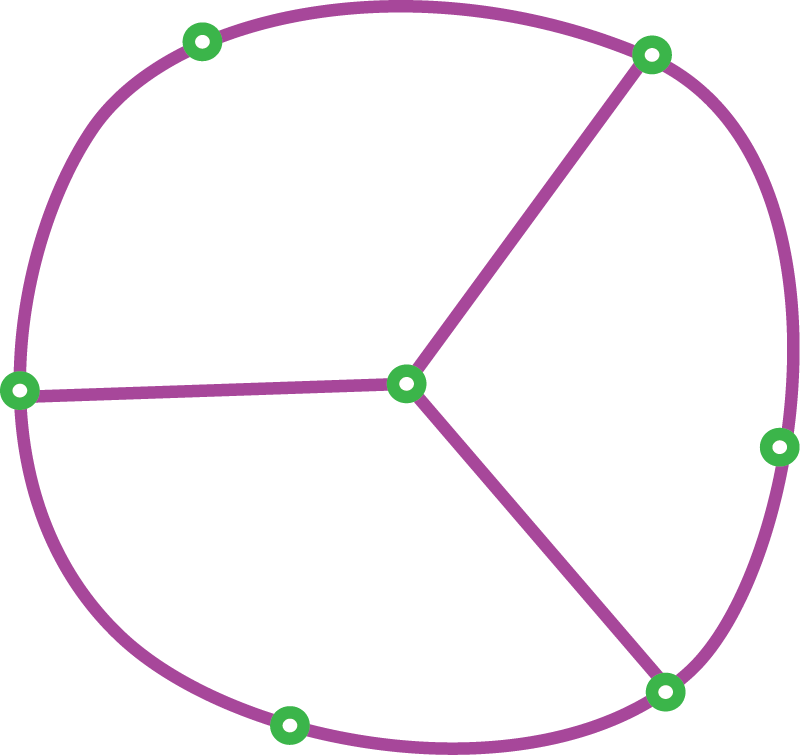}&
\includegraphics[width=0.32\textwidth]{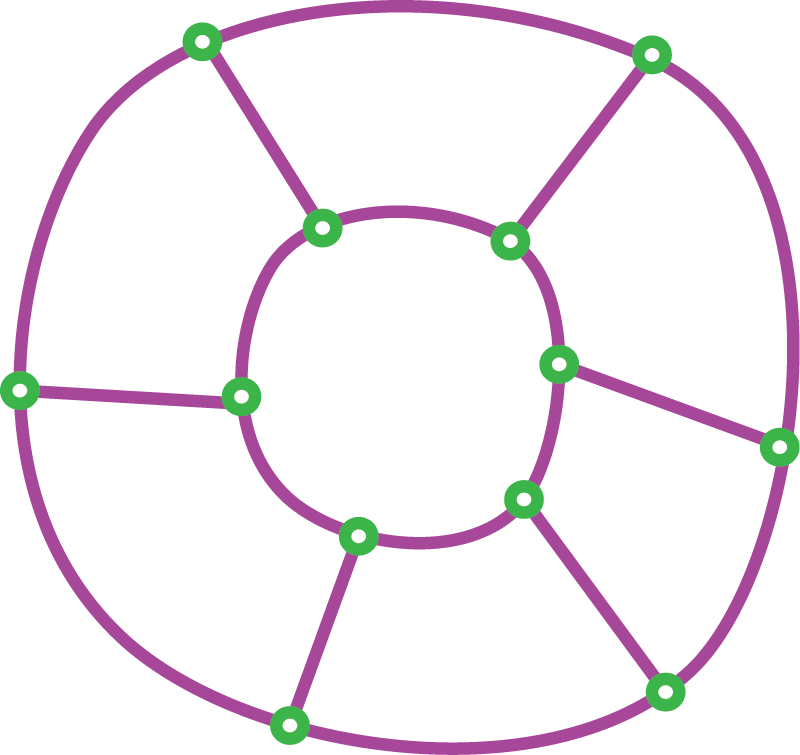}\\
 (a)  Create a grid  &   (b) Create a polygon &   (c) Create a toroid  \\
\end{tabular}
\end{center}
\caption{   Examples of the operations that create quad-heavy meshes.}
 \label{fig_create}
 \end{figure}

\begin{figure}[htbp]
\begin{center}
\includegraphics[width=0.95\textwidth]{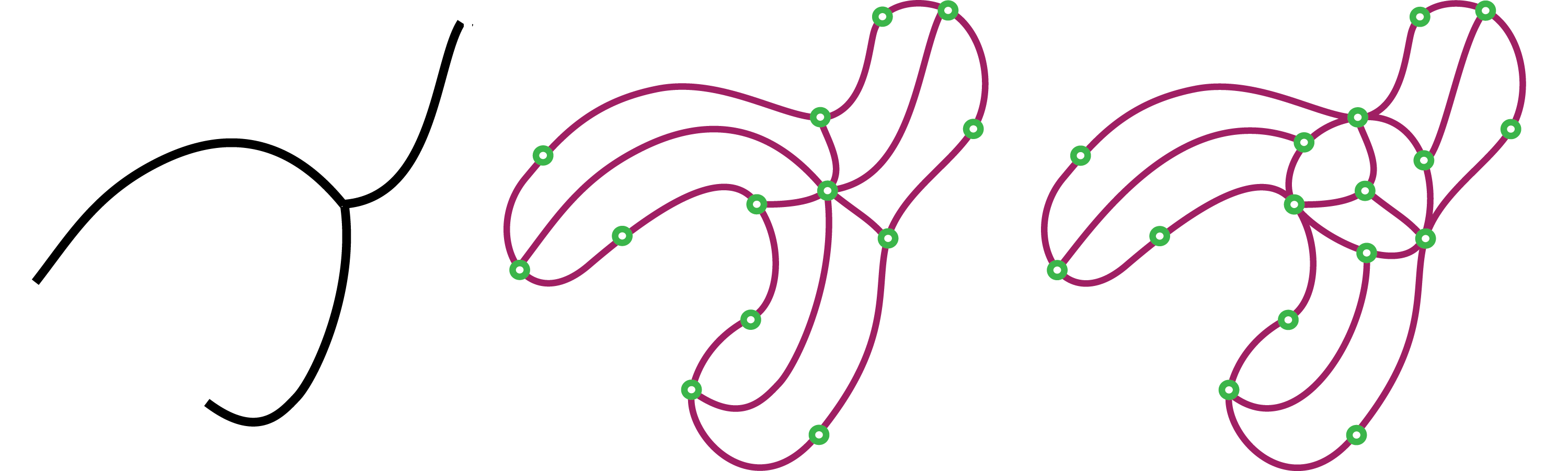}\\
\end{center}
\caption{  Two examples of thickening a medial axis to a quad mesh.}
 \label{fig_create2}
 \end{figure}

\begin{figure}[htbp]
\begin{center}
\includegraphics[width=0.95\textwidth]{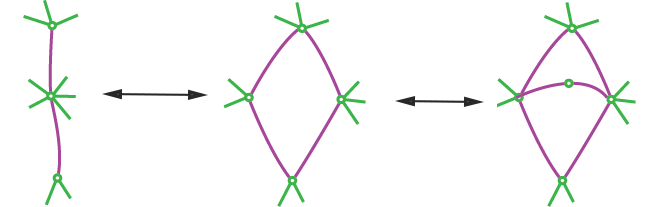}
\end{center}
\caption{   New quad creations with split-edge two-edge and split-face operators.}
 \label{fig_operations}
\end{figure}

Our system allows us to create a set of initial meshes that consists mainly of quadrilaterals. Most of the quadmesh operators created are self-explanatory, as shown in Figure~\ref{fig_create}(b), (c), and (d). The thin medial axis is the only operation conceptually different from the others. In this case, two types of quad mesh are created starting from the user-drawn medial axis, which can be a planar graph created as a network of connected curves, as shown in Figure~\ref{fig_create}(d). Users further manipulate these meshes using quad-preserving operations. We have identified two operations that can introduce a new quadrilateral to the mesh and remove a quadrilateral, as shown in Figure~\ref{fig_operations}(a). We have also generalized existing local operations, such as extruding face and inserting eye operations into group operations, extrude and wrinkle \cite{Akleman2006mod1} (see Figure~\ref{fig_operations}(b)). We have also introduced new local operations, such as insert handle, that can construct a 2-complex (not shown here). These operations are local in the sense that they can be applied in any local area and do not affect the rest. On the other hand, edge split and insert edge operations can only be applied from boundary to boundary. With these operations, the user can create complicated quad mesh structures.

\begin{figure}[htbp]
\begin{center}
\begin{tabular}{cccccc}
\fbox{\includegraphics[height=0.30\textheight]{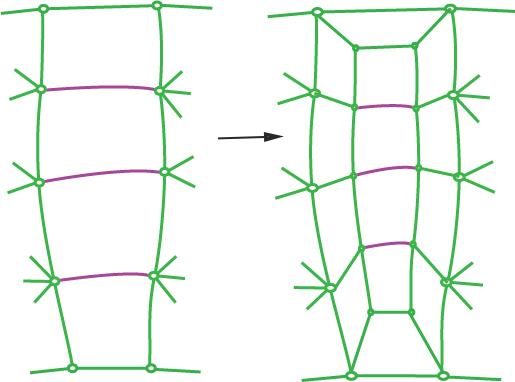}}&
\fbox{\includegraphics[height=0.30\textheight]{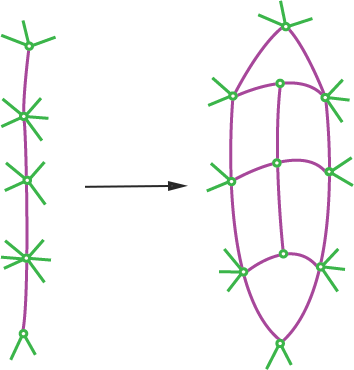}}\\
 (b) extrusion operation & wrinkle operation \\
\end{tabular}
\end{center}
\caption{Two quadrilateral property preserving group operations.}
 \label{fig_operations2}
\end{figure} 

\begin{figure}[ht]
\begin{tabular}{cccccc}
\includegraphics[width=0.32\textwidth]{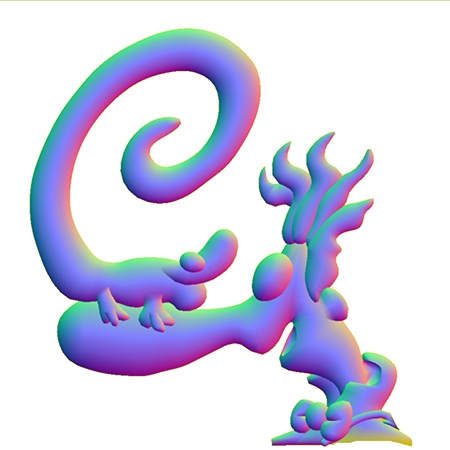}&
\includegraphics[width=0.32\textwidth]{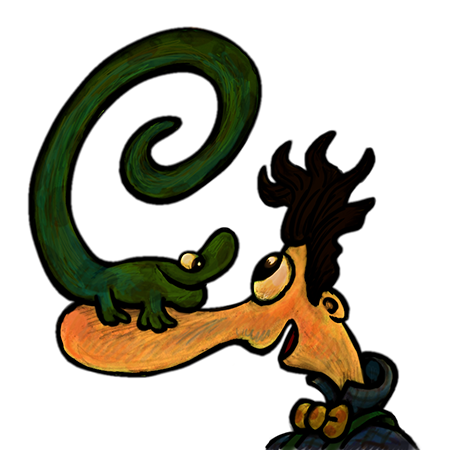}&
\includegraphics[width=0.32\textwidth]{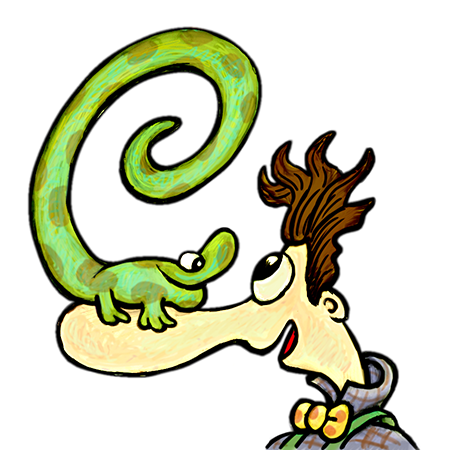}\\
(a) Shape map &
(b) Shader parameter 0 &
(c) Shader parameter 1 &
  \end{tabular}
  \caption{ An example of an object created using our software that turns a static illustration into a dynamic one. (a) a shape map created from the original drawing using our system; (b,c) are the shader parameters.}
\label{fig_bukalemun}
\end{figure}

\begin{figure}[ht]
\begin{center}
\begin{tabular}{cccccc}
 \includegraphics[width=0.45\textwidth]{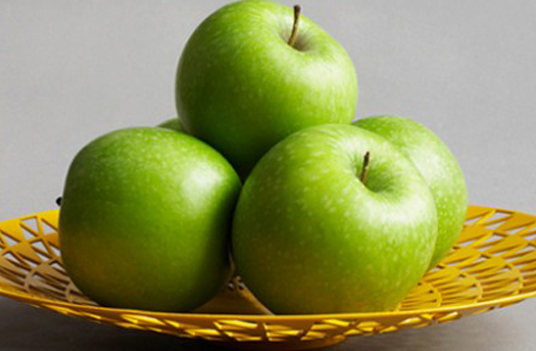}&
 \includegraphics[width=0.45\textwidth]{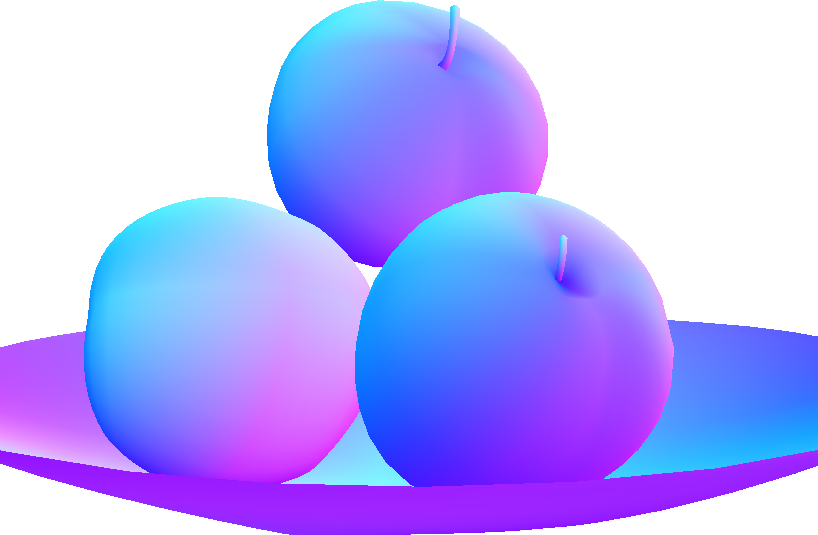}\\
  (a)  Original picture &  (b)  Shape map \\
  \includegraphics[width=0.45\textwidth]{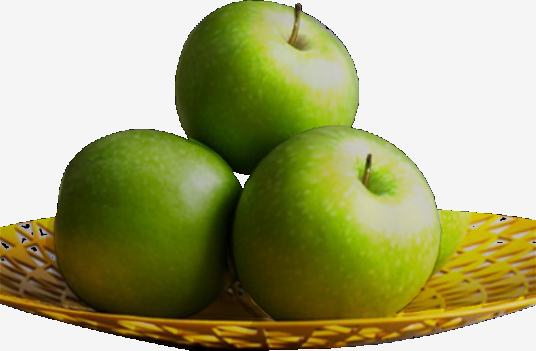}&
 \includegraphics[width=0.45\textwidth]{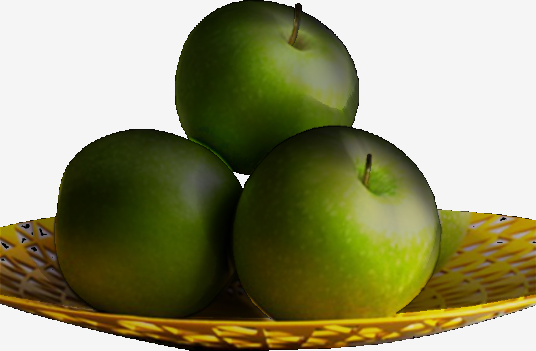}\\
(c)  Image reconstruction &  (d)  Another Re-rendering \\
\end{tabular}
\end{center}
\caption{ Reconstruction apples removing one. }
\label{fig_apple}
\end{figure}

\begin{figure}[ht]
\begin{center}
\begin{tabular}{cccccc}
 \includegraphics[width=0.45\textwidth]{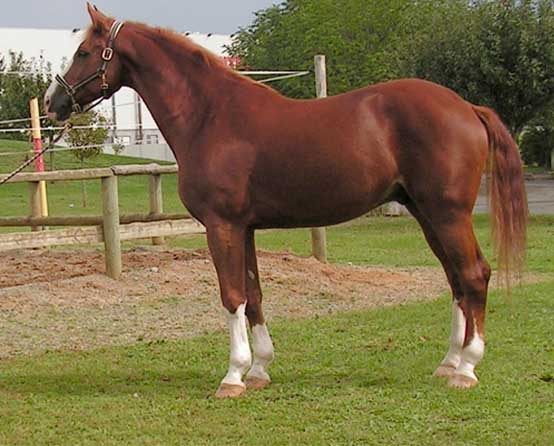}&
 \includegraphics[width=0.45\textwidth]{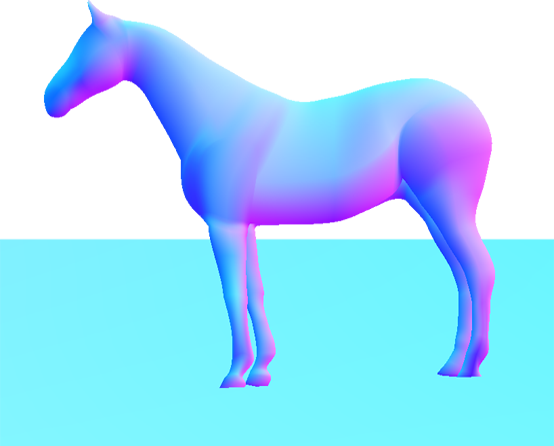}\\
  (a)  Original picture &  (b)  Mock-3D scene \\
  \includegraphics[width=0.45\textwidth]{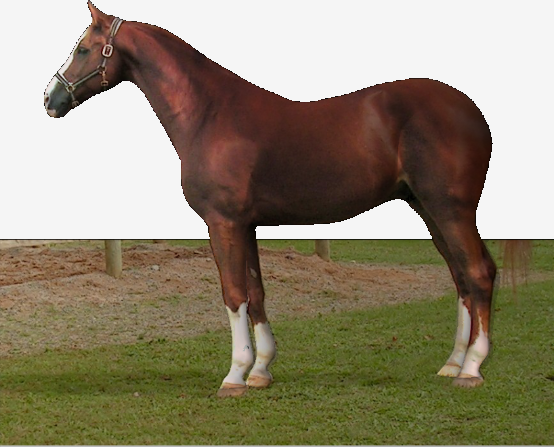}&
 \includegraphics[width=0.45\textwidth]{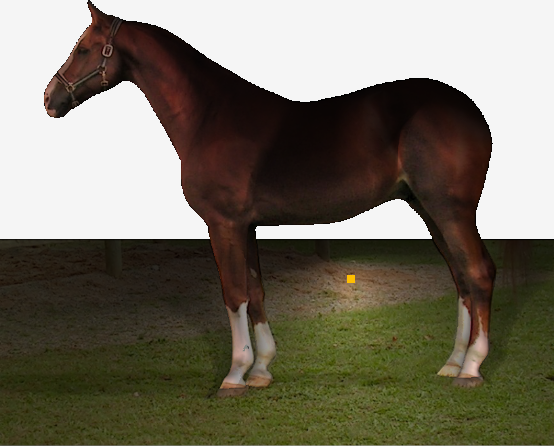}\\
 (c)  Image reconstruction &  (d)  Another Re-rendering \\
\end{tabular}
\end{center}
\caption{ Reconstruction of a horse. }
\label{fig_horse}
\end{figure}

Figures~\ref{fig_vectorfield2},~\ref{fig_homer},~\ref{fig_vectorfield},~\ref{fig_bukalemun},~\ref{fig_apple}, and~\ref{fig_horse} show examples of shape maps created using our prototype system. The mock-3D shape in Figure~\ref{fig_horse} is a 2-complex. Figure~\ref{fig_bukalemun} is an example of a reconstruction of an illustration. Figures~\ref{fig_apple} and~\ref{fig_horse} are examples of photographs' reconstructions. In these two cases, material textures are automatically computed from the photograph using a method similar to single-view relighting \cite{Okabe2006}.  

\section{Conclusion and Future Work}

As stated earlier, we envision a future in which static pictorial documents are converted into dynamic forms that can be accessible and continuously enriched by everyone. To reach this goal, there is a need for the development of (1) a powerful representation that supports general dynamic documents with re-renderable elements and (2) semiautomatic and simple-to-use methods for turning static documents into dynamic documents. In this paper, we have provided a theoretical infrastructure for the development of web-based systems such that, without any additional tool, people can turn their illustrations, artwork, photographs, or cartoons into ``html-like'' documents that can dynamically be rendered, viewed, or manipulated on any device.  

\begin{figure}[ht]
\begin{tabular}{cccccc}
\includegraphics[width=0.32\textwidth]{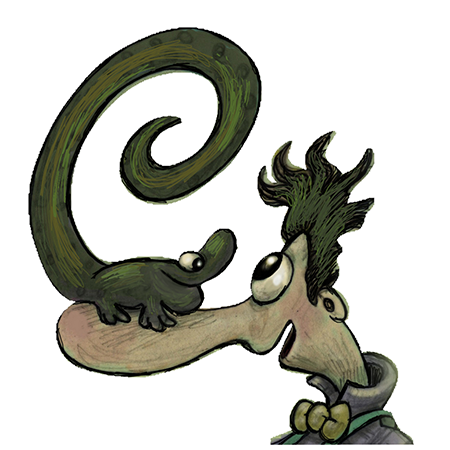}&
\includegraphics[width=0.32\textwidth]{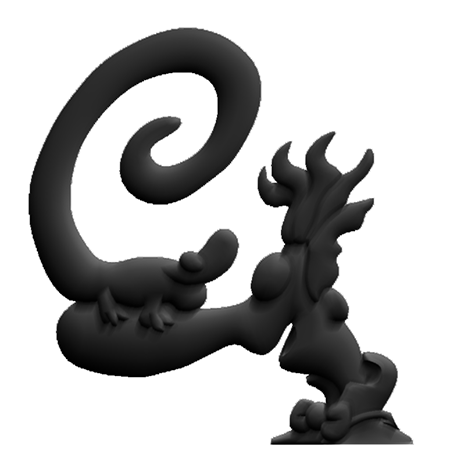}&
\includegraphics[width=0.32\textwidth]{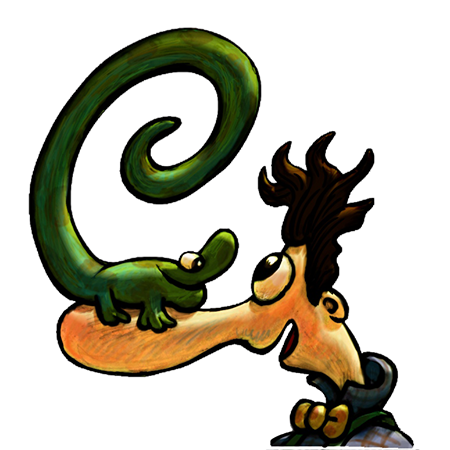}\\
 (a) Original Image&
   (b) B\&W  rendering &
   (c) Color rendering  \\
\end{tabular}
\caption{ (a) An artist's original illustration; (b) a shape map created from the original drawing using our system; (c) a B\&W image that provides the combined effect of our shading, shadow, and ambient occlusion computations; (d) a color-rendering image created by using interpolating texture images. The final image is more volumetric in look than the original drawing due to subtle effects provided by shadow and ambient occlusion even though there is no true 3D shape.}
\label{fig_bukalemun1}
\end{figure} 

Our most important theoretical contribution is to demonstrate that non-conservative vector fields provide a conceptual framework for modeling and representing incoherent, inconsistent, and impossible objects and scenes. We believe that we have presented sufficient evidence, through a variety of examples and methods, to demonstrate the viability of our approach as a non-realistic shape representation. We have implemented a prototype renderer and demonstrated that it is possible to obtain real-time expressive depictions with shadows and global illumination effects using this representation. We also implemented a prototype sketch-based modeling for the interactive construction of mock-3D scenes with 2D  non-conservative vector fields mapped on 2-complexes. Using the system, we have also shown that this approach can be used directly to reconstruct realistic objects and scenes.   

The prototype modeling system provides only proof-of-concept for the theoretical framework. We think that for more general applications, other types of modeling approach, such as implicit and subdivision, should be developed and included in the system. Our current implementation does not support curves, which can be useful for the efficient representation of very thin objects, such as hair. Our general rendering framework can support any of these shape representations since we simply compute geometry in fragment shader by projecting all rays and shapes to the 2D plane.   

In the current framework, we do not allow moving cameras, which is a simple orthographic projection and perspective projection is supposed to be embedded in the mock-3D shapes similar to bas-reliefs. Although we did not implement and/or discuss it, it is possible to make minor rotations without significant visual distortions since the objects already have an embedded thickness associated with them. The current thickness information does not have to be precise to obtain reasonably acceptable shadows and refraction; however, to make extreme rotations such as $90^0$, the thickness information must be carefully created.  For $180^0$ rotations, the back function $F_1$ might need to be created with a separate vector field instead of a single thickness parameter added to $F_0$. Moreover, in extreme rotations embedded perspective information must also be updated. We think that the simplest solution will be to use more than one shape map for visually convincing rotations. A similar approach can also be used to obtain animations (see \cite{Bezerra2005} for a possible direction).  Reconceptualizing non-conservative fields as view-dependent shapes may also be useful for visualization of higher-dimensional vector fields, and it can be an interesting research direction.

\bibliographystyle{unsrtnat}
\bibliography{references}

\end{document}